\documentclass[a4paper]{article}

\usepackage{a4wide}
\usepackage{cite}
\usepackage{amsmath,amssymb,amsfonts}
\usepackage{algorithmic}
\usepackage{graphicx}
\usepackage{textcomp}
\usepackage{bmpsize}
\usepackage{xcolor}
\usepackage{lipsum}
\usepackage[colorlinks=true,urlcolor=black]{hyperref}
\def\BibTeX{{\rm B\kern-.05em{\sc i\kern-.025em b}\kern-.08em
    T\kern-.1667em\lower.7ex\hbox{E}\kern-.125emX}}

\usepackage{booktabs}
\usepackage{comment}
\usepackage[utf8]{inputenc}
\usepackage[T1]{fontenc}
\usepackage{csquotes}
\usepackage{pdflscape} 
\usepackage{makecell}
\usepackage{tabularx}
\usepackage{rotating}
\usepackage{float}
\usepackage{array}
\newcolumntype{x}[1]{>{\centering\arraybackslash}m{#1}}
\usepackage{multirow}
\usepackage{tikz}
\usetikzlibrary{fit, shapes, calc, patterns, patterns.meta}
\usepackage{tikzscale}
\usepackage{wasysym}
\usepackage{fontawesome5}
\usepackage{pifont}
\usepackage[normalem]{ulem} 
\newcommand{\yes}{\CIRCLE}%
\newcommand{\no}{\Circle}%
\newcommand{\half}{\LEFTcircle}%
\newcommand{\rhalf}{\RIGHTcircle}%

\newcommand{\redyes}{\textcolor{red}{\yes}}%
\newcommand{\redno}{\textcolor{red}{\no}}%
\newcommand{\redhalf}{\textcolor{red}{\half}}%
\newcommand{\redrhalf}{\textcolor{red}{\rhalf}}%
\newcommand{\iconPositiveInfo}{positive}
\newcommand{\iconNegativeInfo}{negative}
\newcommand{\iconBothInfo}{both}
\newcommand{\iconTO}{TO}
\newcommand{\iconNTO}{NTO}
\newcommand{\iconOccTO}{occ. TO}
\newcommand{\iconPublicAvailable}{\faGlobe}
\newcommand{\iconPrivate}{\faClipboard[regular]}
\newcommand{\iconAppendOnly}{\faIcon[regular]{clipboard}\raisebox{0.5ex}{\textcolor{teal}{\tiny\faPlus}}}
\newcommand{\iconAsync}{A}
\newcommand{\iconSync}{S}
\newcommand{\iconPartiallySync}{PS}
\newcommand{\iconPublic}{\scalebox{0.7}{\faEye}}
\newcommand{\iconOperator}{\scalebox{0.7}{\faCog}}
\newcommand{\iconUser}{\scalebox{0.7}{\faUser}}
\newcommand{\iconDedicatedEntity}{\scalebox{0.7}{\faSearch}}
\newcommand{\partially}{\LEFTcircle}%
\newcommand{\SP}{\LEFTcircle}%
\newcommand{\SPo}{%
  \hspace{-2pt}%
  \tikz[baseline=-0.6ex]{
    \node[anchor=center, text=yellow] at (0,0) {\rhalf};
    \node[anchor=center, text=black] at (0,0) {\LEFTcircle};
  }\!%
}
\newcommand{\RP}{\RIGHTcircle}%
\newcommand{\iconLive}{L}
\newcommand{\iconSynthetic}{S}

\newcommand{\doublespender}{DS}
\newcommand{\unwanted}{\textcolor{red}{U}}
\newcommand{\iOSS}{\raisebox{-0.12em}{\includegraphics[height=0.75em]{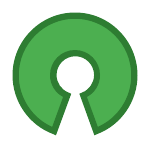}}}%
\newcommand{\numpapers}{36}

\tikzset{>=stealth}
\definecolor{textcolor}{RGB}{62,62,59}
\definecolor{maincolor}{cmyk}{0,0.5,1,0} 
\definecolor{altcolor}{cmyk}{1,0.6,0,0.56} 
\colorlet{mainlight}{maincolor!50}
\colorlet{altlight}{altcolor!50}
\colorlet{textlight}{textcolor!70}
\colorlet{alert}{red}

\usepackage{tikz}
\usepackage{xparse}

\newcommand\addvmargin[1]{
  \node[fit=(current bounding box),inner ysep=#1,inner xsep=0]{};
}

\NewDocumentCommand{\iconCOMMUNICATION}{m m m m m m m O{}}{
    \scalebox{0.4}{
        \begin{tikzpicture}[baseline=(current bounding box.center)]
            \tiny
            \coordinate (O) at (0, 0);
            
            \draw (240:1) node[draw, circle] (S) {};
            \draw (300:1) node[draw, circle] (R) {};

            \ifnum#1=1
                \draw[#8] (S) -- (R);
            \fi

            \ifnum#2=1
                \ifnum#1=6
                    \draw[dashed] (S) -- (O);
                \else
                    \draw[#8] (S) -- (O);
                \fi
            \fi

            \ifnum#3=1
                \ifnum#6=1
                    \draw[dashed] (R) -- (O);
                \else
                    \draw[#8] (R) -- (O);
                \fi
            \fi
            
            \ifnum#4=1
                \draw (0:0.3) node [shape=circle,fill] (O1) {};
                \draw (72:0.3) node [shape=circle,fill] (O2) {};
                \draw (144:0.3) node [shape=circle,fill] (O3) {};
                \draw (216:0.3) node [shape=circle,fill] (O4) {};
                \draw (288:0.3) node [shape=circle,fill] (O5) {};
            
                \ifnum#5=1
                    \draw (O1) -- (O2);
                    \draw (O1) -- (O3);
                    \draw (O1) -- (O4);
                    \draw (O1) -- (O5);
                    \draw (O2) -- (O3);
                    \draw (O2) -- (O4);
                    \draw (O2) -- (O5);
                    \draw (O3) -- (O4);
                    \draw (O3) -- (O5);
                    \draw (O4) -- (O5);
                    \draw (O) -- (O2);
                \else
                    \foreach \x in {O1,O2,O3,O4,O5} {
                        \draw (O) -- (\x);
                    }
                \fi
            \else
                \node[shape=circle,fill] at (O) {};
            \fi
            
            \addvmargin{2mm}
        \end{tikzpicture}
    }
}

\NewDocumentCommand{\devilnode}{m m m O{0}}{
    \ifnum#4=1
        \draw[red, fill=red] (#1) node (#2) {} (#1) circle (0.18);
        \draw[red, fill=red] (#1) ++(65:0.18) -- ($(#1) + (45:0.28)$) -- ($(#1) + (25:0.18)$) -- cycle;
        \draw[red, fill=red] ($(#1) + (155:0.18)$) -- ($(#1) + (135:0.28)$) -- ($(#1) + (115:0.18)$) -- cycle;
    \else
        \ifnum#3=1
            \draw[red, fill=red] (#1) node (#2) {} (#1) circle (0.15);
        \else
            \draw[red] (#1) node (#2) {} (#1) circle (0.15);
        \fi
        \draw[red, fill=red] (#1) ++(70:0.15) -- ($(#1) + (45:0.25)$) -- ($(#1) + (20:0.15)$) -- cycle;
        \draw[red, fill=red] ($(#1) + (160:0.15)$) -- ($(#1) + (135:0.25)$) -- ($(#1) + (110:0.15)$) -- cycle;
    \fi
}

\NewDocumentCommand{\crashnode}{m m m O{0}}{
    \ifnum#4=1
        \draw[orange, fill=orange] (#1) node (#2) {} (#1) circle (0.20);
        \draw[thick, white] (#1) node (#2) {} ($ (#1) + (-240:0.15) $) arc[start angle=-240, end angle=60, radius=0.15];
        \draw[thick, white] (#1) ++ (0, -0.01) --++ (0, 0.17);
    \fi
    \ifnum#4=2
        \draw[orange, fill=orange] (#1) node (#2) {} (#1) circle (0.30);
        \draw[ultra thick, white] (#1) node (#2) {} ($ (#1) + (-240:0.2) $) arc[start angle=-240, end angle=60, radius=0.2];
        \draw[ultra thick, white] (#1) ++ (0, -0.01) --++ (0, 0.23);
    \else
        \ifnum#3=1
            \draw[orange, fill=orange] (#1) node (#2) {} (#1) circle (0.18);
            \draw[thick, white] (#1) node (#2) {} ($ (#1) + (-240:0.12) $) arc[start angle=-240, end angle=60, radius=0.12];
            \draw[thick, white] (#1) ++ (0, -0.01) --++ (0, 0.16);
        \else
            \draw[thick, orange] (#1) node (#2) {} ($ (#1) + (-240:0.15) $) arc[start angle=-240, end angle=60, radius=0.15];
            \draw[thick, orange] (#1) ++ (0, -0.01) --++ (0, 0.17);
        \fi
    \fi
}

\NewDocumentCommand{\iconCENTRALIZED}{O{0}}{
    \scalebox{0.25}{
        \begin{tikzpicture}[baseline=(current bounding box.center)]
            \tiny
            \foreach \angle/\name in {0/N1, 36/N2, 72/N3, 108/N4, 144/N5, 180/N6, 216/N7, 252/N8, 288/N9, 324/N10}
                \draw (\angle:1) node (\name) {} circle (0.15);
            \node (Z) at (0,0) {};
            \begin{scope}
                \foreach \i in {1,...,10}
                    \draw[shorten <=2pt] (N\i) -- (Z);
            \end{scope}
            \ifnum#1=1
                \crashnode{0,0}{Z}{1}[2]
            \else
                \draw[fill] (0,0) node (Z) {} circle (0.3);
            \fi          
            \addvmargin{2mm}
        \end{tikzpicture}
    }
}

\NewDocumentCommand{\iconDECENTRALIZED}{O{0}}{
    \scalebox{0.25}{
        \begin{tikzpicture}[baseline=(current bounding box.center)]
            \tiny
            \foreach \angle/\name in {0/N1, 144/N3, 288/N5} {
                \draw[fill] (\angle:1) node (\name) {} circle (0.15);
            }
            
            \ifnum#1=2
                \foreach \angle/\name in {72/N2, 216/N4} {
                    \devilnode{\angle:1}{\name}{1}
                }
            \else
                \ifnum#1=1
                    \foreach \angle/\name in {72/N2, 216/N4} {
                        \crashnode{\angle:1}{\name}{1}
                    }
                \else
                    \foreach \angle/\name in {72/N2, 216/N4} {
                        \draw[fill] (\angle:1) node (\name) {} circle (0.15);
                    }
                \fi
            \fi
            
            \begin{scope}
            \foreach \a/\b in {
                N2/N1,
                N3/N1, N3/N2,
                N4/N1, N4/N2, N4/N3,
                N5/N1, N5/N2, N5/N3, N5/N4} 
            {
                \draw[shorten <=2pt, shorten >=2pt] (\a) -- (\b);
            }
            \end{scope}
            \addvmargin{2mm}
        \end{tikzpicture}
    }
}

\NewDocumentCommand{\iconFEDERATED}{O{0}}{
    \scalebox{0.25}{
        \begin{tikzpicture}[baseline=(current bounding box.center)]
            \tiny
            \foreach \angle [count=\j from 1] in {0, 51, 102, 153, 204, 255, 306} {
                \draw (\angle:1) node (N\j) {} circle (0.15);
            }
            \draw[fill] (90:.4) node (H1) {} circle (0.18);
            \draw[fill] (210:.2) node (H2) {} circle (0.18);
            
            \ifnum#1=2
                \devilnode{330:.3}{H3}{1}[1]
            \else
                \ifnum#1=1
                    \crashnode{330:.3}{H3}{1}[1]
                \else
                    \draw[fill] (330:.3) node (H3) {} circle (0.18);
                \fi
            \fi
            
            \begin{scope}
                \draw[shorten <=2pt, shorten >=3pt] (N2) -- (H1);
                \draw[shorten <=2pt, shorten >=3pt] (N3) -- (H1);
                \draw[shorten <=2pt, shorten >=3pt] (N4) -- (H2);
                \draw[shorten <=2pt, shorten >=3pt] (N5) -- (H2);
                \draw[shorten <=2pt, shorten >=3pt] (N6) -- (H2);
                \draw[shorten <=2pt, shorten >=3pt] (N7) -- (H3);
                \draw[shorten <=2pt, shorten >=3pt] (N1) -- (H3);
                \draw[thick, shorten <=3pt, shorten >=3pt] (H1)--(H2);\draw[thick, shorten <=3pt, shorten >=3pt] (H2)--(H3);\draw[thick, shorten <=3pt, shorten >=3pt] (H1)--(H3);
            \end{scope}
            \addvmargin{2mm}
        \end{tikzpicture}
    }
}

\pgfdeclarepattern{
    name=my north east,
    parameters={\hatchsize,\hatchlinewidth},
    bottom left={\pgfpoint{-.1pt}{-.1pt}},
    top right={\pgfpoint{\hatchsize+.1pt}{\hatchsize+.1pt}},
    tile size={\pgfpoint{\hatchsize}{\hatchsize}},
    code={
        \pgfsetlinewidth{\hatchlinewidth}
        \pgfpathmoveto{\pgfpoint{-.1pt}{-.1pt}}
        \pgfpathlineto{\pgfpoint{\hatchsize+.1pt}{\hatchsize+.1pt}}
        \pgfusepath{stroke}
    }
}

\tikzset{
    hatch size/.store in=\hatchsize,
    hatch line width/.store in=\hatchlinewidth,
    hatch size=1pt,
    hatch line width=.05pt,
}
\NewDocumentCommand{\share}{m m O{0}}{
    \resizebox{!}{1.5ex}{
        \begin{tikzpicture}[baseline={(current bounding box.center)}, inner sep=0pt, outer sep=0pt]
            \draw[white] (0,-.2) -- (0,1.1);
            \ifnum#3=1
                \fill[red] (0,0) -- (0:1) arc (0:{3.6 * #1}:1) -- cycle;
                \draw[pattern=my north east, pattern color=red] (0, 0) circle (1);
                \fill[white] (0,0) -- (0:1) arc (0:{-3.6 * (100 - #1 - #2)}:1) -- cycle;
                \ifnum\numexpr#1+#2>0\relax
                    \draw[red] (0,0) -- (0:1) arc (0:{-3.6 * (100 - #1 - #2)}:1) -- cycle;
                \fi
                \draw[red] (0, 0) circle (1);
            \else
                \fill[black] (0,0) -- (0:1) arc (0:{3.6 * #1}:1) -- cycle;
                \draw[pattern=my north east] (0, 0) circle (1);
                \fill[white] (0,0) -- (0:1) arc (0:{-3.6 * (100 - #1 - #2)}:1) -- cycle;
                \ifnum\numexpr#1+#2>0\relax
                    \draw[black] (0,0) -- (0:1) arc (0:{-3.6 * (100 - #1 - #2)}:1) -- cycle;
                \fi
                \draw (0, 0) circle (1);
            \fi
        \end{tikzpicture}
    }
}

\title{Systematization of Knowledge: The Design Space of Digital Payment Systems with Potential for CBDC} 

\author{
  \makebox[.2\linewidth]{\normalsize\bfseries Judith Senn}\\
  {\normalfont\itshape\small University of Innsbruck}\\
  {\normalfont\small\texttt{\href{mailto:judith.senn@uibk.ac.at}{judith.senn@uibk.ac.at}}}
  \and
  \makebox[.2\linewidth]{\normalsize\bfseries Aljosha Judmayer}\\
  {\normalfont\itshape\small University of Vienna}\\
  {\normalfont\small\texttt{\href{mailto:aljosha.judmayer@univie.ac.at}{aljosha.judmayer@univie.ac.at}}}
  \and
  \makebox[.2\linewidth]{\normalsize\bfseries Nicholas Stifter}\\
  {\normalfont\itshape\small SBA Research}\\
  {\normalfont\small\texttt{\href{mailto:nstifter@sba-research.org}{nstifter@sba-research.org}}}
  \and
  \makebox[.4\linewidth]{\normalsize\bfseries Rainer Böhme}\\
  {\normalfont\itshape\small University of Innsbruck}\\
  {\normalfont\small\texttt{\href{mailto:rainer.boehme@uibk.ac.at}{rainer.boehme@uibk.ac.at}}}
}
\date{}

\begin{document}

\maketitle

\begin{abstract}

Central Bank Digital Currencies (CBDCs) are proposed as a public response to the uptake of privately run digital payments, with the digital euro, under development by the European Central Bank (ECB), serving as a prominent example.
This momentum provides a unique opportunity to fundamentally rethink the future of money, and, assuming wide adoption, to establish payment systems that offer strong cryptographic security and privacy guarantees from the start.
While the central banks in charge are investigating privacy-enhancing technologies (PETs), they often conclude that PETs are immature or insufficiently scalable.
Moreover, these efforts tend to examine primitives in isolation, offering little insight into how a system using these PETs would scale.

This systematisation of knowledge, therefore, provides a structured, top-down technical analysis of \numpapers{} payment system designs of complete system proposals that can inform CBDC designs or were explicitly proposed for this application.
We identify recurring design patterns, technical trade-offs, and implementation challenges. 
Concluding, we highlight research gaps, including offline payments and post-quantum security.

\end{abstract}
\section{Introduction}
Money is vital for the economy. 
With the rise of online payments, it is increasingly handled by private firms. 
Central banks have become alert to this trend, as shown by their reaction to Facebook's initiative to launch Libra~\cite{merch2019frankfurt}.
Payment revenues account for about 2 percent of gross domestic product in the US, 3 percent in Asia, and 1 percent in the EU, largely driven by fees~\cite{de2021pay}. 
The growing dominance of the private sector also implies lost seigniorage and limited monetary policy options~\cite{cipollone2025ljubljana}.
This is reinforced by the decline of cash~\cite{bis2025paymentstats}, and the expanding popularity of cryptocurrencies and stablecoins~\cite{kosse2023making}.
In response, central banks have started to explore how the public sector can issue money in digital form~\cite{aurazo2024central,hub2020projectHelvetia}, leading to numerous initiatives to research or pilot Central Bank Digital Currencies (CBDCs)~\cite{cbdctracker}.
To do so, some central banks have hired cryptographers and formed tech teams. 
However, it is not easy for them to ramp up their technological expertise to match that in their core domains of finance and economics, where their staff are often at the forefront of research.

The design of a retail payment system involves three objectives that are inherently in tension with one another: privacy, compliance, and performance.
Privacy is essential because payment data can reveal sensitive personal information~\cite{de2015unique,auer_mapping_2023}.
At the same time, privacy cannot be unconditional, since this would facilitate misuse such as money laundering~\cite{fatf2012recommendations}.
This makes technical support for compliance essential.
Adding to the challenge is the need for scalability, as a retail CBDC may need to process hundreds of thousands of transactions per second with low latency~\cite{boj_cbdc_performance_2024,ecb_digital_euro_market_research_2023}.
Existing digital payment systems prioritise reliability, scalability, and compliance. 
Technical privacy protections are rare~\cite{garratt2021privacy,preibusch2016shopping}. 
Introducing a new digital retail payment system gives central banks the opportunity to reshape digital money with stronger security and privacy guarantees.
Achieving this, however, is far from easy.

To address these challenges, several central bank studies~\cite{arora2025privacy,torres2024privacy,asrow2021privacy,stella_phase4,auer_PETS_2025} as well as academic works~\cite{DBLP:conf/aft/BaumCDF23} survey Privacy Enhancing Technologies (PETs) as potential building blocks for a CBDC.
While these studies offer valuable insights, the bottom-up approach is problematic because PETs cannot be composed arbitrarily. 
As a result, these studies provide limited understanding of how systems incorporating PETs would perform and scale.
This calls for a top-down perspective that compares complete system proposals. 
Some authors have made progress in this direction.
Nardelli et al.~\cite{nardelli2025hitchhiker} present a broad comparison, but they analyse privacy without taking performance into account.
Chatzigiannis et al.~\cite{chatzigiannis2021sok} emphasise compliance and privacy, but do not propose a comprehensive framework. 
Other works discuss economic implications of privacy~\cite{agur2022designing,auer2021central} or general privacy--compliance trade-offs~\cite{allen_design_2020,kiff2020survey}.
All of these works are worth reading and have inspired our systematisation.

To the best of our knowledge, the present work is the most comprehensive technical top-down study.
We analyse a large set of systems, with a consistent taxonomy and an emphasis on designs with potential for CBDC.
Our main contributions are as follows:
\begin{itemize}
    \item We develop a framework to compare system architecture choices, privacy, compliance, performance, and other technical features with implications for the user experience (such as offline functionality), capturing 28 properties, including assumptions and undesired properties.
    \item We compare \numpapers{} payment system designs, covering several decades of research, from coin-based systems to distributed ledgers. 
    Our categorisation allows us to compare these vastly different designs, reaching from Chaumian e-cash, over selected cryptocurrencies, to recent CBDC proposals.
    \item We reduce complexity by mapping the design space into four characteristic patterns and discuss characteristic features of each.
    \item We highlight underexplored areas and open challenges, such as post-quantum security and benchmarking methods reflecting real-world transaction behaviour.
\end{itemize}
Although the title of this work refers to digital payment systems with potential for CBDC, our paper selection is not limited to explicit CBDC proposals.
We deliberately adopt a broader scope for two reasons.
First, making informed decisions about the design of a CBDC requires an understanding of what is technically possible. 
Much of the recent progress in digital payments has been made in the cryptocurrency space.
Second, there is no sharp technical boundary between CBDCs and cryptocurrencies.
Neither term has a universally accepted technical definition.
Although both support digital payments, they are typically based on very different trust assumptions: CBDCs assume a central authority at least for issuance, while most cryptocurrencies aim to circumvent any authority. 
These assumptions influence the resulting system designs. 
However, they are insufficient to create a clear distinction between the two.
Which design is most suitable for building a CBDC ultimately depends on the relevant policy objectives.

This paper is organised as follows:
Section~\ref{sec:money} sets the stage with a brief elaboration on the concept of money.
Section~\ref{sec:categories} introduces the framework for comparing payment designs.
Section~\ref{sec:results} presents the results of our analysis in two tables and outlines four design patterns.
Section~\ref{sec:discussion} discusses selected properties of the design patterns. 
Section~\ref{sec:conclusion} concludes.
The exact methodology for paper selection is detailed in Appendix~\ref{xsec:method}.
For context, we assume that readers are familiar with the current banking system~\cite{bech2020innovations} and have basic knowledge of Bitcoin~\cite{Nak08} and Chaumian e-cash~\cite{Cha82}.
\newpage
\section{What is Money? What is a Payment?}
\label{sec:money}
 
There is no universal agreement on the definition of money, even amongst economists:
``\emph{few issues in economics have generated such heated debates as the nature of money and its role in the economy.}''~\cite[p.~1]{borio2019money}.  
Money is the distribution of value combined with social conventions that enable its exchange for goods and services~\cite{de2021pay}.
The success of money depends on public trust.  
In the case of central bank money, this trust relates to the central bank's capacity to act in the public interest~\cite{bis2025ar}.   
Economists tend to define money by the functions it fulfils: unit of account, means of payment, and store of value~\cite{hicks1967CriticalEssays}.  
Closer to the mindset of computer scientists is Kocherlakota's famous argument that money is memory~\cite{kocherlakota1998money}, more precisely an \emph{imperfect} representation of records of all past transactions.
The latter perspectives will guide our discussion.  

One function of money is to serve as a means of payment.  
To pay, the sender must \emph{authorise} the payment, and it must be ensured that only they can do so.  
The recipient, in turn, must be able to \emph{validate} the correctness of the payment.
Otherwise, they cannot rely on the social convention that the value can be exchanged in the future.
In this way, money combines individual (authorisation) and social (validation) interests.
For cash, ownership of a banknote authorises the holder to spend it, and validation occurs by inspecting the physical properties of the banknote that make it difficult to reproduce. 

In the case of digital money, this logic must be mapped to memory, i.\,e., information preserved over time.
The sender holds private information (e.g., secret keys) that enable them to authorise payments.  
We refer to information held by these parties as \emph{local information}.  
In practice, users can outsource local information to trusted intermediaries who can interact with the payment system on their behalf.
By contrast, the information required to validate transactions is referred to as \emph{global information}.
It may include records of previous transactions, the distribution of value in the system,
as well as public keys. 
Global information is essential for detecting and preventing the same units from being spent more than once (double-spending). 
This ensures that digital money cannot be easily copied.

We define a \emph{system operator} as an entity responsible for maintaining the global information.
Operators can be distributed and may include central banks, commercial banks, other payment service providers, or miners. 
Global information may be public (as in a ledger) or restricted to privileged operators.
In conventional online banking, local information consists of a user's authentication credentials and account identifiers, while global information is kept  by banks in the form of account balances and transaction records.
In Bitcoin, local information is represented by private keys, and global information is maintained in a distributed ledger recording unspent transaction outputs (UTXOs).
In Chaumian e-cash, local information takes the form of anonymous tokens signed by the issuer, whereas the database of spent tokens used to detect double-spending represents the global information.

The system state is defined by the combination of all local and global information. 
It may not be fully observable by any single party.
Value transfers imply updates of the system state.
System architectures differ in how they represent value in the system state. 
They also differ how state information is distributed locally and globally.
All this has implications on performance, the achievable level of privacy, options to enforce compliance, and usability.  

\begin{figure}[t]
    \centering
    \includegraphics[width=\columnwidth]{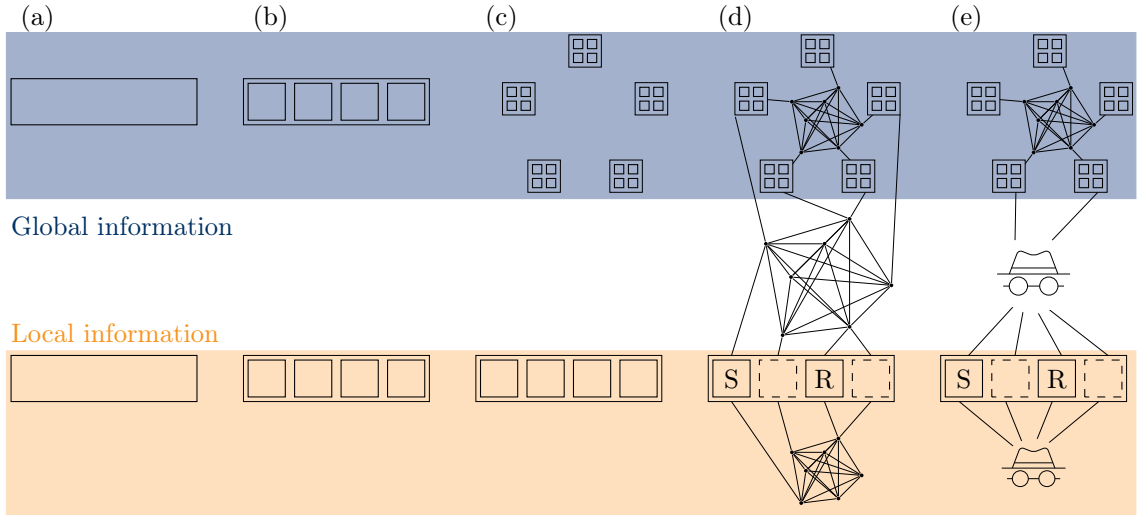}
    \caption{
    Stepwise illustration of core concepts:
    (a) local and global information;
    (b) distribution of local and global information;
    (c) replication of global information;
    (d) communication networks between users, between users and operators, and inter-operator communication;
    (e) anonymous communication channels to ensure privacy against observers of the network traffic.
    }
    \label{fig:information}
\end{figure}

Illustrating this in a diagram, every digital payment system involves local information used to authorise payments and global information used to validate them (see Fig.~\ref{fig:information}a).  
Local information must be distributed among users, each holding the information required to authorise their own funds.  
Global information can be distributed among multiple operators, such as commercial banks maintaining different user accounts (see Fig.~\ref{fig:information}b).  
Moreover, global information is often replicated to ensure robustness against crashes and data loss (see Fig.~\ref{fig:information}c).  
However, one important aspect of payments remains to be discussed.  
A payment requires communication between the sender and the recipient, as well as between the users and the system operators, and among the operators themselves (see Fig.~\ref{fig:information}d).
The privacy of all these communication channels, as well as the communication patterns, must be considered when evaluating a system.

According to Kahn, money is not only memory, but also privacy~\cite{kahn_2005_money_is_privacy}.  
He argues that the value of money may lie in its imperfect memory -- in other words, in the presence of privacy, understood as missing information.  
Too much information in a payment system might not only lead to general harms such as chilling effects, censorship, and suppression, but also specific economic consequences.  
It could enable price discrimination~\cite{Hannak2014PriceDiscrimination,aquistini2006conditioningPrices}.  
Additionally, a lack of privacy may cause money to lose its fungibility, as different funds may carry different associated information. 
In the worst case, money might even cease to function as an accurate unit of account~\cite{moeser2014towards}. 
If every payment involves a bundle of value and information, whose exact value is difficult to determine, it becomes impossible to measure the value of the goods being exchanged.  
Finally, since payments are a fundamental component of all digital economic interactions in our society, the privacy provided by the payment infrastructure effectively bounds the level of privacy that can be achieved. 
In other words, if payments lack privacy, any interaction involving payments will, too.
Such a lack of privacy in payments risks desensitising society, normalising surveillance, and lowering expectations of privacy in other areas of life.
Moreover, from a data minimisation perspective~\cite{EU2016GDPR}, it is also desirable to only process and store the minimal information required for payment processing to reduce the impact of data breaches.

For all these reasons, public retail payment systems should safeguard privacy.  
Technically, this requires two steps:
First, the communication flows in the network must be anonymised. 
It is not sufficient to hide the sender and recipient identifiers if their communication remains visible 
(see Fig.~\ref{fig:information}e).  
Second, the amount of potentially compromising global information must be minimised.

Through the lens of privacy, it is instructive to distinguish two ways of representing global information.
The system may record what can be spent (\emph{positive information}) or what has already been spent and thus must not be spent again (\emph{negative information}).  
An example of positive information is account balances that are updated with each transaction, implying that the transaction graph is observable.  
In contrast, negative information can take the form of nullifiers stored whenever a value is spent, preventing double-spending (e.g., Chaumian e-cash~\cite{Cha85}).  
In this case, the transaction graph---except for pathological cases---remains hidden, but the list of nullifiers grows indefinitely. 
Reducing global information to the point where the transaction graph can no longer be reconstructed often means relying on negative information.\footnote{
   Exceptions exist, e.g. transactions in zkLedger~\cite{NVV18} update all positive global information via Pedersen commitments and zk-proofs.
}
\section{Technical Design Categories} 
\label{sec:categories}

In this section, we describe the framework that defines the lens through which we compare the payment designs proposed in the literature and the columns of our comparison tables \ref{tbl:main1} and \ref{tbl:main2}.
We can broadly classify the categories into 
\emph{system architecture} (Section~\ref{ssec:architecture}), 
\emph{assumptions and trust} (Section~\ref{ssec:assumptions}), 
\emph{privacy} (Section~\ref{ssec:privacy}), 
\emph{compliance} (Section~\ref{ssec:compliance}), 
\emph{performance} (Section~\ref{ssec:performance}), and
\emph{technical features with implications for the user experience} (Section~\ref{ssec:tech_feat}).
We discuss the classes and their categories in this order, noting that some forward references are unavoidable as many properties are interdependent.

% =====================================================================================
\subsection{System Architecture} 
\label{ssec:architecture}

In this category, we capture the main design choices. 
Since terms such as \emph{account model}, \emph{UTXO model}, and \emph{token} are often used ambiguously and inconsistently in the literature, we introduce novel criteria for system comparison.\footnote{
    The notions of tokens and accounts are highly overloaded and have been defined in different ways. 
    For example, a working paper of a central bank~\cite{CGM21} classifies Bitcoin as account-based in order to position e-cash as token-based. 
    By contrast, the cryptocurrency community distinguishes between UTXO and account models~\cite{buterin2016thoughts, zahnentferner2018chimeric, ACD19}, while referring to accounts managed by smart contracts as ``tokens''~\cite{eip20,wang_tokenisation_on_blockchain}. 
    The confusion increases further when hardware tokens as authentication devices are included in the discussion. 
    In card payment systems, ``tokenization'' has yet another meaning, namely the practice of hiding account identifiers from merchants~\cite{bis2020tokenization}.
}
In Section~\ref{sssec:information}, we revisit our classification of global information into positive and negative information.
Section~\ref{sssec:integrity} discusses how the integrity of this global information is ensured, while Section~\ref{sssec:visibility} examines who can access or observe it.
We then address the communication required to perform a payment (Section~\ref{sssec:communication}) and, finally, the distributed systems aspects of the design (Section~\ref{sssec:network}).

\subsubsection{Information}
\label{sssec:information}

In Section~\ref{sec:money} we introduced the terms \emph{local information}, needed to authorise a payment, and \emph{global information}, required to validate it.
Recall that we can further classify global information into:
\begin{enumerate}
\item \emph{Positive information}, which allows the system to verify the existence of value.
If value is spent, the corresponding entry of positive information must be updated or deleted.
\item \emph{Negative information}, which allows the system to detect the non-existence of value.
If value is spent, negative information must be added.
Negative information is not updated or deleted.
\end{enumerate}
Positive information can be more compact, but comparing differences between states may reveal the transaction graph, i.\,e., who has paid whom.
Negative information does not reveal the transaction graph, but has the downside that the amount of global information grows forever. 
Systems can combine both types of global information.

\subsubsection{Integrity of Global Information} 
\label{sssec:integrity}

Many systems distribute the global information across several entities and infrastructures.
As information propagates through the network at different speeds and across varying distances, transient faults can cause entities to hold inconsistent views of the global information.
Such discrepancies can enable \emph{double-spending attacks}, where an adversary exploits stale or inconsistent system states to deceive a victim into accepting a transaction that later proves invalid~\cite{stifter2022opportunistic}.
To mitigate these inconsistencies, systems differ in how they coordinate and, in particular, order transactions across entities.
We distinguish:

\begin{enumerate}
    \item \emph{Total order} (TO), where all operators \emph{reach agreement} on the same order of transactions, i.e., they are linearised into a consistent log.
    This requires relying on a centralised system operator or distributed computing primitives such as \emph{consensus}~\cite{pease1980reaching} or \emph{atomic broadcast}/\emph{total order broadcast}~\cite{cristian1995atomic}. 
    We do not distinguish whether the agreement on a total order is reached deterministically or eventually, and whether the mechanism is supported with incentives.\footnote{
        We refer the reader to \cite{bano2019sok, garay2020sok} for further details on consensus.
    }
    \item \emph{No total order} (NTO), where the ordering requirement for transactions is relaxed and does not require a global total ordering.\footnote{
        This is in particular applicable to asset transfer, which facilitates concurrent processing. 
        See~\cite{sridhar2025stingray} for cases where total order is required.
    }
    In this setting, double-spent funds---whether due to error or malicious intent---become unspendable~\cite{sliwinski_abc_2020}. 
    Under the assumption that participants prefer not to lose funds, this is an incentive mechanism that discourages double-spending.
\end{enumerate}
Total order can provide strong consistency guarantees and prevent double-spending; however, it generally incurs communication and computational overhead. 
Recent works~\cite{sridhar2025stingray,blackshear2024sui} propose a \emph{delayed total order} approach, where agreement is periodically invoked after an epoch.
This combines the throughput of NTO with the stronger guarantees of TO, ensuring that no funds are permanently lost.
For example, PEReDi~\cite{KKS22} follows this approach by invoking Byzantine agreement when a user requests an abort. 
In Table~\ref{tbl:main1}, this is denoted with \iconOccTO{}.

Independent of the total order property, global information can be \emph{sharded} horizontally to improve performance.
Sharding introduces additional integrity requirements and can increase protocol complexity.
Further discussion is provided in Appendix~\ref{xsec:sharding}.

\subsubsection{Visibility of Global Information and History}
\label{sssec:visibility}

Systems differ in how widely the state is shared and whether historical states are retained.
We distinguish:
\begin{enumerate}
\item \emph{Private} (\iconPrivate) refers to a distributed database in which only operators have read access to individual records and can update them as specified in the previous subsection.
Not all operators may have access to all records.
Historical records are not required for the operation, and it remains at the discretion of the operators to delete them.
\item \emph{Private, append-only} (\iconAppendOnly) refers to the special case of a database distributed between the operators, where transactions can be appended, but existing records are never modified.
Append only databases are often called ``ledgers''\footnote{In the literature, ledgers are often implicitly assumed to be append-only. We hereby explicitly imply ledgers to be append-only.}. 
There is no expectation that historical records are deleted.
\item \emph{Public} (\iconPublicAvailable) means that anyone can receive all state updates. 
Since deletion of information cannot be enforced for an open set of participants, this implies that the history of all transactions is public, but not necessarily in plaintext. From a \emph{visibility} perspective, we assume that public global information is implicitly append-only. 
\end{enumerate}

\subsubsection{Communication}
\label{sssec:communication}
Communication is a fundamental aspect of any payment system.
Different use cases may favour different communication relations. 
While sender and recipient may interact at the point of sale, this is impractical for asynchronous payments, such as invoices.

We represent communication types using triangular icons, where the bottom-left node denotes the sender, the bottom-right the recipient, and the top the operator.
Lines between nodes imply that communication on this relation is necessary to process a payment.
Lines between nodes indicate mandatory communication links; dashed lines indicate deferrable communication (b).
Icons (c) and (d) illustrate distributed operators, with (d) requiring inter-operator communication.
In these cases, direct communication with the recipient is not required to update the global information. 
Nevertheless, the recipient must access the global information in order to obtain the updated state.

\begin{center}
(a)~\iconCOMMUNICATION{1}{0}{1}{0}{0}{0}{0}\hspace{1em}
(b)~\iconCOMMUNICATION{1}{0}{1}{0}{0}{1}{0}\hspace{1em}
(c)~\iconCOMMUNICATION{0}{1}{0}{1}{0}{0}{0}\hspace{1em}
(d)~\iconCOMMUNICATION{0}{1}{0}{1}{1}{0}{0}
\end{center}

Examples of these communication types are provided in Appendix~\ref{xsec:communication}.
\subsubsection{Network and Distributed System Model}
\label{sssec:network} 
Payment systems are inherently distributed.
Communication delays and component failures can compromise the integrity of global information, which makes it essential to understand the environment in which the payment system operates as intended, referred to as \emph{the distributed system model}.
This model specifies the assumed properties of the communication network and the failure behaviour of system components and entities (see Section~\ref{sssec:trust_operator}).

When designing and reasoning about distributed systems, abstractions and simplifications are necessary to model their components and the real-world environment in which they operate.
These abstractions help focus on the core problems that require formal solutions~\cite{lynch1996distributed}.
For instance, a payment system design may assume reliable communication channels or allow users to behave arbitrarily.
These assumptions determine which fundamental distributed computing problems are solvable and which are not.
It is therefore essential that papers define their system model.
Many proposed designs omit or insufficiently describe their system model, which can cause discrepancies between design and implementation and lead to vulnerabilities or loss of desired properties.
For this reason, we record for each design the level of detail provided about the \emph{distributed system model}: 
\no{} indicates no description, 
\partially{} a basic model, and 
\yes{} a detailed one.

Second, we analyse the network model for each of the payment designs.  
First, we determine which model of synchrony the inter-operator communication assumes.
\begin{enumerate}
    \item \emph{Synchronous (\iconSync):}  
    There is a known upper bound for the network delay that always holds. 
    \item \emph{Partially synchronous (\iconPartiallySync):}  
    Assumption of a global stabilisation time (GST), after which message delays remain bounded~\cite{dwork1988consensus}.
    \item \emph{Asynchronous (\iconAsync):}  
    There can be unbounded delay between sending and receiving a message.
\end{enumerate}

Further, we evaluate whether any of the communications necessary for a payment require \emph{anonymous} communication to preserve the level of privacy promised by the design.
Here we distinguish: (a) anonymous communication between the users (sender and the recipient) is necessary, (b) anonymous communication between users and operators is necessary, or (c) anonymous communication between both the users themselves and between users and operators is necessary.

\begin{center}
(a)~\iconCOMMUNICATION{1}{0}{0}{0}{0}{0}{0}[dotted, thick] \hspace{1em}
(b)~\iconCOMMUNICATION{0}{1}{1}{0}{0}{0}{0}[dotted, thick] \hspace{1em}
(c)~\iconCOMMUNICATION{1}{1}{1}{0}{0}{0}{0}[dotted, thick]
\end{center}

% =====================================================================================
\subsection{Assumptions and Trust}
\label{ssec:assumptions}

No practical system can be secure against all conceivable adversaries. 
Meaningful security guarantees require clearly stated assumptions defining the conditions under which they hold~\cite[Chapter~1.4]{katz2007introduction}. 
Many payment systems rely on a broad range of assumptions.
These include mathematical and specifically cryptographic assumptions, defining which computations are infeasible, such as the discrete logarithm or Diffie--Hellman assumptions.
Systems may also rely on physical assumptions that reflect physical limitations, such as the non-existence of a practical quantum computer or specific hardware limitations.
Behavioural assumptions specify how parties, such as operators and users, are expected to act.
Commonly, this includes a subset of operators required to be honest and follow the protocol.

Relying on strong assumptions can simplify or even render a design possible; however ensuring that these guarantees hold can be difficult to achieve in practice.
In order to relax the assumptions, additional measures can be employed. 
For example, Byzantine fault tolerance allows weakening honesty assumptions, such that a bounded subset of operators may behave arbitrarily. 
While these measures can allow a system to operate in more hostile environments, they may negatively impact other properties, mainly performance.

We therefore compare payment designs by the assumptions they rely on.
Specifically, we selected assumptions that impact performance.
First, we discuss reliance on a \emph{trusted setup} (Section~\ref{sssec:trusted_setup}), which may improve performance since many efficient zero-knowledge proof systems depend on it.
Next, we examine \emph{trusted hardware} (Section~\ref{sssec:TH}), which can improve performance by replacing cryptographic operations.
Finally, we analyse behavioural assumptions related to system operation (Section~\ref{sssec:trust_operator}), issuance (Section~\ref{sssec:trust_issuer}), and privacy revocation (Section~\ref{sssec:trust_revocation}).
Stronger behavioural assumptions can improve efficiency, while mechanisms to tolerate misbehaviour may reduce it.

\subsubsection{Security Proof and Type of Evidence} 
\label{sssec:proof}

This category indicates whether a paper offers a security proof for at least one claimed property of the payment system.
The mark \yes{} designates that a proof is present, but it does not mean that we have verified it.
We further record the \emph{proof paradigm} by classifying the type of evidence presented:
\begin{enumerate}
    \item \emph{Proof \textbf{sketch}}, a weak form of evidence that requires further analysis;
    \item \emph{Proof-of-concept (\textbf{PoC})}, demonstrating functional aspects of the design\footnote{A proof-of-concept typically verifies the capabilities and operations a design provides (functional properties), while omitting most non-functional properties, such as security, privacy, or reliability~\cite{robertson2012mastering}.};
    \item \emph{A \textbf{deployed} system} has been tested in the real-world. 
    It provides empirical evidence of functionality and some security assurance.
    \item \emph{\textbf{Game} and simulation based proofs}, typically encompassing selected aspects of the design; and
    \item \emph{Universal composability (\textbf{UC})} arguments showing that the proposed design can be derived from an ideal functionality in several steps that are indistinguishable for the adversary~\cite{canetti2020universally}.\footnote{
        This type of evidence includes multi-protocol UC~\cite{ACD19} and approaches inspired by UC~\cite{TBA22}.
    }
\end{enumerate}
Providing a proof means that the authors have made efforts to design the system according to clearly defined security goals, such as realising an ideal functionality. 
However, it does not necessarily translate into real-world assurance of a system's overall security. 
This disconnect has sparked an ongoing debate within the security community about the role and limits of formal proofs in practice~\cite{herley2017sok}. 
None of the designs reviewed here employ machine-verifiable proofs---despite cryptographic library developers increasingly promoting them as a means of improving assurance and auditability~\cite{barbosa2021sok}.
This highlights that there is still a long way to go for payment protocols to enjoy the same scrutiny as other core internet protocols, such as Transport Layer Security (TLS)~\cite{DBLP:conf/snapl/BhargavanBDFHHI17,DBLP:conf/ccs/CremersHHSM17,DBLP:conf/sp/BhargavanBK17,DBLP:journals/popets/ArfaouiBFNO19,RFC8446}.

\subsubsection{Trusted Setup} 
\label{sssec:trusted_setup}
A \emph{trusted setup} refers to a cryptographic setup phase required for secure operation.
It typically involves generating public parameters from private random inputs that must be deleted afterwards.
This solely refers to the setup phase and does not make any statement about trusted parties involved in the operation of the payment system. 

An example for a trusted setup is Zcash~\cite{BCG14}.
Its zk-SNARKs require a one-time setup to compute public parameters ($pp$) from confidential inputs.
If these inputs are not deleted, the holder could forge proofs. 
In Zcash, this would allow undetectable creation of money.
Zcash mitigates this risk through multi-party computation (MPC) protocols~\cite{DBLP:conf/sp/Ben-SassonC0TV15,DBLP:journals/iacr/BoweGM17,DBLP:conf/fc/BoweGG18} that succeed if at least one of several parties is honest and deletes their input.
MPC protocols offer a well-established technique to distribute trust at the expense of higher procedural complexity.\footnote{See Wang et al.~\cite{DBLP:journals/iacr/WangCB25} for an overview of trusted setup protocols and recent so-called ``ceremonies.''}

If a design relies on a trusted setup, we mark it with \redyes{}, and with \redno{} otherwise.
We classify only \emph{private coin} setups~\cite{DBLP:journals/iacr/WangCB25} as trusted.
\emph{Public coin}\footnote{Also referred to as \emph{transparent setups}~\cite{DBLP:conf/eurocrypt/BunzFS20}.} setups, such as those in Pedersen commitments~\cite{DBLP:conf/crypto/Pedersen91}, rely on public randomness (e.\,g., a random beacon) and do not require secret inputs. 
Between otherwise equivalent systems, the one without a trusted setup is preferable.

\subsubsection{Trusted Hardware} 
\label{sssec:TH}

Some designs depend on secure elements, such as tamper-resistant smart cards~\cite{rankl2004smart}, Hardware Security Modules (HSMs)~\cite{sommerhalder2023hardware}, Trusted Platform Modules (TPMs) \cite{tomlinson2017introduction,TCG_TPM2_Library}, or Trusted Execution Environments (TEEs)~\cite{sabt2015trusted}. 
The security of these \emph{trusted hardware} (TH) components depends not only on the hardware's own tamper-resistance, but also on the behaviour of the manufacturer and the supply chain. 

We collect whether trusted hardware is required \redyes{} or not \redno{}. 
Systems that require trusted hardware but have a fallback mechanism to detect circumvention of the trusted hardware are marked with \redhalf{}.
For example, PayOff~\cite{BZK24} enables the retrospective detection of double-spending during the deferred settlement of offline payments.

\subsubsection{Trust in the Operator}
\label{sssec:trust_operator}

The \emph{operator} of a payment system is responsible for maintaining the state of the global information.
This involves verifying payment requests and updating the global information accordingly, although parts of the operation can be delegated to other parties.
Since the operator ultimately decides the global state, it has final authority over the distribution of value in a payment system.
That is why assumptions on the behaviour of the operator are essential when designing a payment system.

Many designs introduce mechanisms to reduce the level of trust required in the operator. 
This can be achieved by
(i)~by distributing the control over the operation,
(ii)~by increasing the tolerance against (arbitrary) faults, or
(iii)~by ensuring that the actions of the operator(s) are auditable.

In some systems, the operator's role is distinct from that of the \emph{issuer}, or from entities responsible for compliance and privacy revocation.
The same three approaches to reduce trust apply to these roles as well, and the icons introduced below are used consistently across roles.

\begin{table}[ht]
    \centering
    \caption{Two dimensions of trust in the operator(s)}
    \begin{tabular}{lccc}
    \toprule
    & \multicolumn{3}{c}{\textbf{Fault tolerance}} \\
                       \cmidrule{2-4}
                       \textbf{Distribution}
                        & None 
                        & Crash                 
                        & Byzantine             \\
	\midrule                        
        Centralised     & \iconCENTRALIZED   &                       &                       \\
        Federated       & \iconFEDERATED     & \iconFEDERATED[1]     & \iconFEDERATED[2]     \\
        Decentralised   & \iconDECENTRALIZED & \iconDECENTRALIZED[1] & \iconDECENTRALIZED[2] \\
       \bottomrule
    \end{tabular}
    \label{tab:operator_assumptions}
\end{table}

\noindent
\textbf{Distribution of Control.}
Control can be distributed to reduce reliance on a single trusted party.
We distinguish:
\begin{enumerate}
    \item \emph{Centralised}: a single trusted entity has final authority (Table~\ref{tab:operator_assumptions}, row~1);
    \item \emph{Federated}: a fixed set of known operators collectively govern state updates (row~2);
    \item \emph{Decentralised}: all participants have equal privileges, and there is no distinction between users and operators (row~3).
\end{enumerate}

\noindent
\textbf{Fault Tolerance.}
Distribution alone does not ensure resilience against failures.
We distinguish:
\begin{enumerate}
    \item \emph{Crash fault tolerance}: the system keeps operating correctly even if a bounded subset of operators becomes unavailable (Table~\ref{tab:operator_assumptions}, column~2);
    \item \emph{Byzantine fault tolerance}: the system keeps operating correctly even if a bounded subset of operators acts arbitrarily, which includes intentionally malicious actions (column~3).\footnote{
        In this paper, as well as in the literature, the terms \emph{malicious} and \emph{corrupt} are used interchangeably. 
        }
\end{enumerate}
Fault tolerance increases resilience by ensuring correct system operation despite absent or subversive operators.
This can be relevant for currency areas such as the eurozone, where national central banks may act as operators, but not all of them are entirely free from political influence.
A single, centralised operator cannot be fault tolerant itself; however, some components of the system 
may still be, as demonstrated in Project Hamilton~\cite{LVF23}.

\vspace{\baselineskip}
\noindent
\textbf{Auditability of System Integrity.}
Auditability, in this context, refers to the capability to verify that the system operates according to the predefined rules. 
This includes ensuring that no valid transactions are omitted without justification.
Depending on the type of actors who can audit the system, we distinguish whether \emph{any party} can verify the correct operation (\iconPublic) or whether only the system \emph{operator(s)} can verify the correct operation (\iconOperator).
Although auditability does not guarantee correct operation, it reduces the necessary trust, since potential misbehaviour can be detected.

\subsubsection{Trust in the Issuer}
\label{sssec:trust_issuer}

While in most cases the operator(s) are also responsible for issuing currency units, it is also possible to assign the role of issuing new currency units to a dedicated party, the \emph{issuer}. 
The issuer controls the creation of value within the system and thus determines the overall supply.
This role naturally aligns with a central bank in the context of a CBDC.
The level of trust in the issuer can be reduced using the same principles described for the operator in Section~\ref{sssec:trust_operator}.
We again use the icons from Table~\ref{tab:operator_assumptions} to represent the distribution of control and fault tolerance.
For issuance, \emph{auditability} refers to the ability to verify new issuance of value and to observe the total supply.
This improves transparency and prevents unnoticed inflation. 
Auditing can be performed by the public (\iconPublic), the system operator(s) (\iconOperator), or a dedicated entity (\iconDedicatedEntity).

\subsubsection{Trust in the Privacy Revocation}
\label{sssec:trust_revocation}

Privacy is an important aspect of any CBDC and the respective properties will be described in Section~\ref{ssec:privacy}.
Some systems allow user privacy to be revoked in the event of suspicious behaviour.
We refer to the entity responsible for this process as the \emph{privacy revocation authority}.
Trust in this authority can be reduced using the same principles applied to the operator and the issuer.
We again use the icons from Table~\ref{tab:operator_assumptions} to indicate the distribution of control and fault tolerance.
Auditability of privacy revocation refers to the ability to detect when a user's privacy has been revoked.
It helps to hold the authority accountable and detect disproportionate surveillance.
With \iconUser{} we indicate that the affected user can detect this action.
Note that privacy revocation in response to double-spending is not considered here, as such revocation can be necessary to keep the integrity of the system state.
For designs that do not offer any privacy revocation, this column is left empty.

% =====================================================================================
\subsection{Privacy} 
\label{ssec:privacy}

Privacy is a central concern in the design of a payment system.
Laws such as the GDPR~\cite{EU2016GDPR} in Europe regulate the collection, processing and transfer of personal data, which includes payment records.
While privacy is a fundamental right that protects individuals from unwarranted surveillance, it also limits the ability to detect and prosecute fraud, illicit payments, and money laundering.
To balance these opposing needs, payment system designs have historically proposed varying degrees of privacy.

It is convenient to measure the privacy properties of a payment system by standardised notions.
For example, Auer et al.~\cite{auer_mapping_2023} distinguish between \emph{anonymity} (e.\,g., \emph{someone} receives 100\,€ from \emph{somebody}), \emph{pseudonymity} (e.\,g., pseudonym$_A$ receives 100\,€ from pseudonym$_B$), and \emph{confidentiality} (e.\,g., Alice receives a transaction from Bob), or combinations thereof. 
While these notions are intuitive for a general audience, they are insufficient to fully assess privacy because it is not  a property of a single transaction in isolation. 
As information accumulates over time, a set of transactions which initially appears anonymous can later become pseudonymous. 
The linkability of Monero's ring signatures is an example of this~\cite{MoserSHLHSHHMNC18}.
To address information accumulation, we follow Pfitzmann and Hansen's~\cite{pfitzmann2010terminology} terminology and adapt it to payments.

The strongest privacy notion is \textbf{unlinkability}. 
In \cite[p.~12]{pfitzmann2010terminology}, it is defined as:
``\emph{Unlinkability of two or more items of interest (IOIs) from an attacker's perspective means that within the system, the attacker cannot sufficiently distinguish whether these IOIs are related or not.}'' 
For our purposes, we take transactions and real-world user identities (short: identities) as the items of interest and define unlinkability in the context of payments as follows:
Unlinkability of two or more \textit{transactions or identities} means that an attacker cannot determine whether these \textit{transactions or identities} are related.\footnote{See~\cite{DBLP:conf/pet/SteinbrecherK03,DBLP:journals/csur/WagnerE18} for formal treatments of unlinkability.}

Unlinkability may not be achievable due to constraints in the logic of payments.
Imagine a system with two wallets, one of which is empty. 
The empty wallet can never be the sender of a transaction. 
Moreover, as payment systems update state sequentially, temporal relations cannot always be hidden.
Therefore, we use notions of partial unlinkability to capture the nuances of existing designs.

\subsubsection{Identity Privacy}
\label{sssec:identity_privacy}
A system offers \emph{identity privacy} if---except for pathological cases---observation of the system does not reveal significant information about the relation between a real-world identity and any party involved in a transaction. 
If this privacy property is met for both the sender and the recipient, we denote it by \yes. 
If this property is fulfilled only for the sender, we denote it by \SP, and if only for the recipient, by \RP.
Otherwise, we denote it by \no.
The special case in which identity privacy is ensured for the sender, while for the recipient it is guaranteed only in the case of an offline payment, is denoted by \SPo.  
This corresponds to the identity privacy in a world where bank transfers and cash coexist. 
Identity privacy corresponds to a notion of pseudonymity offered by cryptocurrencies that do not identify parties but reveal the transaction graph, such as Bitcoin and Ethereum.

\subsubsection{Value Privacy}
\label{sssec:value_privacy}
A system offers \emph{value privacy}\footnote{Other terms  in the literature capturing value privacy are \textit{transaction privacy}, \textit{confidentiality}, or \textit{transaction amount obfuscation}.} if---except for pathological cases---observation of the system does not reveal significant information about the relation between the transaction values of any two transactions (e.\,g., whether the value of transaction $A$ is the same, higher, or lower than the value of transaction $B$).
If this property is met, we denote it by \yes. 
If the value of the transaction is not disclosed, but transactions can be set into a relation based on their values, we denote it as \partially. 
Otherwise, we denote it by \no.

\subsubsection{User Unlinkability}
\label{sssec:user_unlinkability}
A system offers \emph{user unlinkability} if it cannot be determined---except for pathological cases---whether the same user is involved in two or more transactions as sender or recipient. 
If the anonymity set is the entire set of privacy-preserving transactions, then this privacy property is met, and we denote it by \yes. 
If unlinkability holds only within a subset of parties involved in the transaction, we refer to it as \emph{user subset unlinkability} and denote it as \half. 
Otherwise, we denote it by \no.
Some payment designs specify user unlinkability as the desired level of privacy, as linking transactions can reveal payment patterns and, if a user is deanonymised once, many of their transactions may be linked.  
However, this level of privacy makes compliance checks more challenging.  

\subsubsection{Sender Anonymity}
\label{sssec:sender_anon}
Before, we have introduced the concept of identity privacy.
This can be achieved in two ways.
First, users may have long-term identifiers that are not linked to their real-world identity.
Second, senders remain unknown by cryptographic means, such as in Chaumian e-cash~\cite{Cha82}.
To distinguish these cases, we introduce the notion of \emph{sender anonymity}.
A system provides sender anonymity if the sender cannot be determined. 
If this property holds, we denote it by \yes.
If the sender cannot be determined within a defined subset, we denote it by \half.
If no sender anonymity is provided, we denote it by \no.
This is a weaker form of user unlinkability.
In the literature, sender anonymity is also referred to as unlinkability~\cite{CHL05}.
Their terminology collides with ours.

While theoretically possible, we do not consider recipient anonymity as a category. 
All systems that offer user unlinkability also provide recipient anonymity, making the distinction unnecessary.

\subsubsection{Adversary}
To evaluate whether any of these privacy properties hold, we consider a computationally bounded adversary, specifically a probabilistic polynomial-time (PPT) adversary, capable of corrupting any party except the sender, the recipient, and the authority responsible for privacy revocation (if this role exists).
Collusion is allowed up to a number smaller than the threshold required for collaborative deanonymisation~\cite{DBLP:conf/fc/KellerFB21}, which we consider a feature rather than a weakness.
If a design supports multiple privacy levels, we map the strongest possible.

% =====================================================================================
\subsection{Compliance}
\label{ssec:compliance}

Every payment system must comply with the law.
The applicable laws and contractual obligations for payment system providers are vast, vary between jurisdictions, and become even more complicated when payments cross borders~\cite{Bech2020}. 
While several designs make efforts to enable compliance, their features are justified with abstract assumptions about what the law might require.
The justification often refers to the \textbf{AML} (Anti-Money Laundering) and \textbf{CFT} (Countering the Financing of Terrorism) regulations implemented internationally after the 9/11 attacks~\cite{Pie02}. 
The \textbf{KYC} (Know Your Customer) principle is a common measure to support effective AML and CFT by holding banks responsible for having identifiers of account holders.
The triple of AML, CFT, and KYC is presumably the greatest common denominator across jurisdictions, but by no means complete.
For example, the regulations of consumer protection, data protection, fraud prevention, tax enforcement and financial stability vary by country or economic area, and affect the payment system.

For the purpose of this systematisation, we distinguish whether the compliance measures apply to the user or the operator. 
In this subsection, we focus entirely on the approaches taken to ensure user compliance.
Other mechanisms address operator-related compliance, such as the distribution of control, which reduces the power of malicious operators, and auditability, which enables misbehaviour to be detected (see Section~\ref{sssec:trust_operator}).
Privacy measures can also protect users from operators and ensure that personal data is handled lawfully (Section~\ref{ssec:privacy}).

We further distinguish broadly between two main approaches to limit undesired behaviour: 
(1) detecting non-compliance after the fact  in order to create a credible threat of sanction (see Section~\ref{sssec:detect_and_sanction}), and 
(2) preventing non-compliance through constraints built into the system (see Section~\ref{sssec:prevent}).

\begin{figure}[t]
    \centering
    \includegraphics[width=\columnwidth]{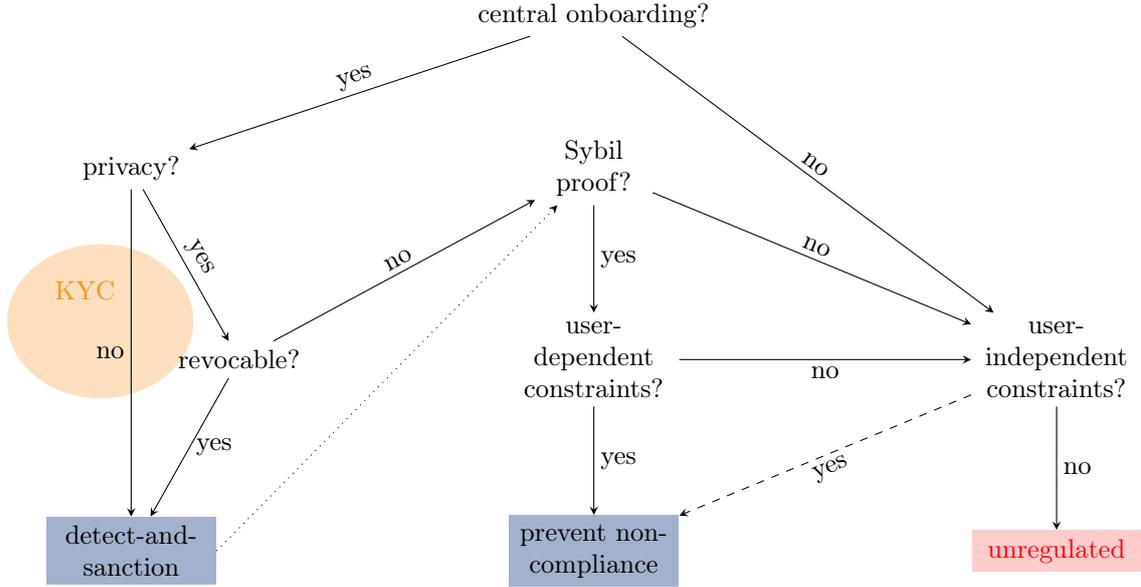}
    \caption{Compliance approaches for digital payment systems based on central onboarding, Sybil resistance, limits, and privacy. 
    The dotted arrow indicates that both approaches can be combined. 
    The dashed arrow indicates a weaker form of preventing non-compliance.
    }
    \label{fig:regulation}
\end{figure}

Fig.~\ref{fig:regulation} visualises the feasibility of the two compliance approaches as a function of system features.
Branches ending at \textcolor{red}{unregulated} without passing the detect-and-sanction box are not desirable, and pose a risk of widespread misuse.
In some systems, different components may follow distinct approaches; for example, Chaumian e-cash implements detect-and-sanction for recipients but imposes no compliance requirements on senders.
Consequently, it is insufficient to classify a system by a single compliance type.

\subsubsection{Detection of Non-Compliant Users} 
\label{sssec:detect_and_sanction}
The detect-and-sanction approach relies on retrospective transaction analysis and escalation through legal processes.
To function effectively, it requires a central onboarding, either without privacy (KYC) or with revocable privacy to enable reliable user identification.
A credible threat of sanction may deter abuse.
Privacy-preserving systems can include a mechanism for \emph{privacy revocation}, where privacy can be lifted for the purpose of detecting non-compliance (e.g. PEReDi~\cite{KKS22}, PayOff~\cite{BZK24}).

That is why we assess whether a system allows the revocation of user privacy.
We distinguish between \yes{} if revocation is possible, \no{} if it is not, \textbf{\doublespender{}} if it occurs only in cases of double-spending, and \textbf{\unwanted{}} if the privacy of honest users may be unintentionally revoked as a side effect of detecting a double-spender.
Beyond double-spending cases, privacy revocation generally implies the (potentially threshold-based) encryption of sensitive information that an authorised auditing party can decrypt for investigations.

\subsubsection{Preventing Non-Compliant User Behaviour}
\label{sssec:prevent}
The prevention-based approach seeks to make non-compliance technically infeasible or highly impractical.
Systems following this model often impose limits on user activity, such as transaction limits, spending limits, or maximum account balances.
It is possible to enforce complex regulatory constraints without unnecessarily reducing privacy by using zero-knowledge proofs~\cite{GGM16, CMT20}. 
However, these methods remain difficult to deploy in practice because they are prohibitively slow with well-established cryptography.
Faster variants are rather new and arguably have not passed the test of time.
Systems enforcing user-specific constraints require both central onboarding and resistance to \emph{Sybils}~\cite{douceur2002sybil}.
Otherwise, users could evade restrictions by creating multiple identities (Sybil attack).
By contrast, global constraints (e.\,g., deny lists) do not require any central onboarding.
In practice, limits rarely eliminate misuse entirely, but they do reduce its potential scope.
This applies, for example, to value or transaction limits, which reduce the amount that can be misused.

We record whether a system can enforce sending (S), receiving (R), transaction (T), or holding (H) limits. 
An asterisk ($*$) denotes limits based on predefined budgets rather than range proofs.

% =====================================================================================
\subsection{Performance}
\label{ssec:performance}

This section defines categories to evaluate performances. 
First, we focus on the availability of an implementation (Section~\ref{sssec:implementation}), and benchmark results (Section~\ref{sssec:benchmarks}).  
Then we highlight the necessity for making payments sequentially as a limiting factor for performance (Section~\ref{sssec:concurrency}).

\subsubsection{Availability of Implementations}
\label{sssec:implementation}
Implementations are absolutely necessary to evaluate performance.
If the described protocol has been fully implemented, we indicate this with \yes{}.  
If only parts of the protocol have been implemented, we use \partially{}.  
If no implementation exists, we mark it with \no{}.  
If the implementation is released under an open-source license, we add the open-source icon (\iOSS{}).

\subsubsection{Benchmark Results}
\label{sssec:benchmarks}
If available, we report the order of magnitude of the claimed \emph{throughput} (Tx/s) and \emph{latency}.
Throughput is the measure of how many transactions a system can process in one second.
Latency denotes the time until a transaction is considered confirmed (for probabilistic designs such as Bitcoin, we use measurements reported in the literature).
We take benchmarks corresponding to the highest privacy level, in line with our privacy evaluation.
When no real-world deployment exists, we rely on the reported claims.
This is why we distinguish:
\begin{itemize}
    \item \emph{Synthetic benchmarks} (S): Based on simulated workloads approximating expected throughput and latency under typical conditions.
    Extreme scenarios, such as bulk salary payments, are often omitted, which can lead to overestimated performance.
    \item \emph{Live benchmarks} (L): Based on real-world deployments reporting throughput and latency under actual operational conditions.
\end{itemize}
As the figures for synthetic benchmarks are taken directly from the original sources and are not normalised, we report only rough orders of magnitude.
This is relevant since testing conditions might differ substantially with respect to hardware, infrastructure, workload, and stress test assumptions. 
As the reported figures span several decades of research, the evolution of hardware capabilities cannot be ignored.
Therefore, the benchmarks only serve as an indication of possible scaling regimes.

\subsubsection{Sequential Payments}
\label{sssec:concurrency}
Allowing a user to be involved in more than one payment without involving an operator after each, is not only a usability concern but also a key factor influencing system performance.
The restriction on being involved in payments sequentially cannot be resolved simply by common scalability techniques, such as distributing the operator's workload horizontally.
Therefore, we record this constraint as a potential bottleneck.
We distinguish: 
both sender and recipient can participate in only one transaction at a time (\redyes), 
only the sender is constrained (\redhalf), 
only the recipient is constrained (\redrhalf), 
payments do not need to be sequential (\redno{}).

\subsubsection{Global Information Growth}
\label{sssec:storage_size_growth}
The chosen type of global information affects scalability, particularly storage size and search efficiency.
Designs based on negative global information require an ever-growing list, making searches increasingly time consuming as the list expands.
Such lists are limited by the available memory and the computational resources needed for searching and inserting entries.
Periodic resets become necessary.
However, they are non-trivial, as they add overhead and affect privacy and usability.
To improve search efficiency, data structures, such as Bloom~\cite{DBLP:journals/cacm/Bloom70}, Cuckoo~\cite{DBLP:conf/conext/FanAKM14}, Vacuum~\cite{DBLP:journals/pvldb/WangZS019} or Quotient filters~\cite{DBLP:conf/sigmod/PandeyCDBFJ21} can be used.
Other solutions involve cleverly choosing the anonymity set such that negative information can be deleted under defined conditions~\cite{DBLP:journals/popets/CachinW25}.
Each providing different trade-offs in space and performance.
At an estimated global rate of 110~000 Tx/s~\cite{bruno2024global}, a Bloom filter that is reset annually would require about 20 TB of memory, illustrating the engineering challenges of nullifier-based designs (see Appendix~\ref{xsec:bloom} for details on our estimation).

That is why we capture whether a design has infinitely growing global information (\redyes), 
or not (\redno{}).
Note that some designs that use positive information for the global state, such as blockchains, never delete information and therefore have an ever-growing global state, too.

% =====================================================================================
\subsection{Technical features with implications for the user experience} 
\label{ssec:tech_feat}

Usability is a key factor in the adoption of payment systems~\cite{MALLAT2007413}, and studies show that it can also influence security~\cite{DBLP:conf/IEEEares/KaindaFR10}.
Most surveyed papers, however, pay little attention to usability, reflecting the early or prototypical stage of many designs.
Since explicit usability considerations are largely absent from the surveyed papers, we instead focus on technical features that may affect usability at later stages of design development.
We focus on the technical features  inherent to the payment design itself---specifically, whether offline payments are possible (Section~\ref{sssec:offline}), whether the system allows for a single payment of exact value (Section~\ref{sssec:exact}), and whether recipient involvement is required (Section~\ref{sssec:interactive}).
User interface quality, key or password recovery, and similar features are excluded, because such aspects can be improved with supporting infrastructure.
The properties discussed here are intrinsic to the payment design and cannot easily be added later.

\subsubsection{Offline Payment}
\label{sssec:offline}

This category captures whether a payment system supports offline functionality, that is, whether payments can be made without communicating with the system operator.
A key attribute of cash is the ability to make payments offline.  
A digital version of offline money could preserve some of the benefits of cash without the physical handling costs.  
Offline functionality offers several advantages.  
It increases resilience by allowing payments during network or power outages~\cite{de2018societal}, supports financial inclusion in regions with poor connectivity, and can enhance privacy.
For instance, the ECB suggests that offline digital euro payments could offer cash-like privacy guarantees~\cite{ECB2025faq}.
At the same time, such functionality might add to the challenges for compliance.
We distinguish three cases: 
full offline functionality (\yes), 
no offline functionality (\no), 
and one-time offline functionality (\partially).
In the latter, payments occur without operator involvement, but the recipient must interact with the operator to reuse or exchange the received funds, an approach taken by early e-cash designs~\cite{CFN88}.

\subsubsection{Paying Exact Amounts}
\label{sssec:exact}
This category captures whether any amount can be transferred exactly in a single payment.
We indicate this property as either supported (\yes) or not supported (\no).
If a system supports payments of varying amounts but only through multiple transactions (fewer than in the trivial case of, e.\,g., issuing 100 units as $100 \times 1$) or if the system restricts payments to specific amounts (e.\,g., powers of two), we denote the property as \partially.

\subsubsection{Uninvolved Recipient}
\label{sssec:interactive}

We evaluate whether the recipient can remain uninvolved in the payment process (\yes) or must actively participate (\no).
If the recipient must provide input that only they can compute and that differs for each payment, they are considered to be involved.
Communicating an address or identifier once does not make the recipient involved.  
If the recipient can remain uninvolved, the system behaves similarly to push payments, such as wire transfers. 
\section{Results}
\label{sec:results}

Tables~\ref{tbl:main1} and \ref{tbl:main2} document the results of our literature analysis.
We constructed them by applying the categories from the previous section to classify \numpapers{} designs for payment systems (See Appendix~\ref{xsec:method} for details on the paper selection).
Besides comparing individual designs, the tables enable us to identify four design patterns, presented in Section~\ref{ssec:data_patterns}.
Section~\ref{ssec:pattern_quant} provides a quantitative overview of the analysed designs per pattern with respect to privacy, usability, and performance properties.

\renewcommand{\arraystretch}{1} 
\begin{table*}[!htpb]  
  \centering
  \caption[caption]{
    Part 1 of the classification of \numpapers{} designs according to Section~\ref{sec:categories} sorted by year of publication. Undesirable properties are shown in red.
  }
  \resizebox{\linewidth}{!}{
  \begin{tabular}{l|l||c|c|l|c|c|c|c||c|c|c|c|l|l|l||c}

\multicolumn{2}{c||}{\textbf{General info}} &
\multicolumn{7}{c||}{\textbf{System architecture}} &
\multicolumn{7}{c||}{\textbf{Assumptions and trust}} &
\\

Year &
Reference &
\makecell{Global \\ information} &
\makecell{Integrity \\ (global)} &
\makecell{Visibility \\ (global)} &
\makecell{Comm. \\ pattern} &
\makecell{Dist. s. \\ model } &
\makecell{Sync. \\ model} &
\makecell{Anon. \\ network} &
\makecell{Security \\ proof} &
\makecell{Type of \\ evidence} &
\makecell{Trusted \\ setup} &
\makecell{Trusted \\ hardware} &
\makecell{Operation} &
\makecell{Issuance} &
\makecell{Privacy \\ revocation} &
\makecell{Design \\ pattern} \\ \hline

1982 & \cite{Cha82} (Blind Sig.) & \iconNegativeInfo & \textcolor{black}{\iconTO} & \iconAppendOnly & \iconCOMMUNICATION{1}{0}{1}{0}{0}{0}{0} & \no &  & \iconCOMMUNICATION{1}{0}{0}{0}{0}{0}{0}[dotted, thick] & \no & \makecell{PoC} & \redno & \redno & \iconCENTRALIZED[0] ~ \hspace*{-5ex} ~ \iconOperator & \iconCENTRALIZED[0] ~ \hspace*{-5ex} ~ \iconOperator & \iconCENTRALIZED[0] & \makecell{Transfer\\ Digital Coin}
            \\ \hline 
1988 & \cite{CFN88} (E-Cash) & \iconNegativeInfo & \textcolor{black}{\iconTO} & \iconAppendOnly & \iconCOMMUNICATION{1}{0}{1}{0}{0}{0}{0} & \no &  & \iconCOMMUNICATION{1}{0}{0}{0}{0}{0}{0}[dotted, thick] & \no & \makecell{PoC} & \redno & \redno & \iconCENTRALIZED[0] ~ \hspace*{-5ex} ~ \iconOperator & \iconCENTRALIZED[0] ~ \hspace*{-5ex} ~ \iconOperator &  & \makecell{Transfer\\ Digital Coin}
            \\ \hline 
1991 & \cite{OO91} (Off-line E-Cash) & \iconNegativeInfo & \textcolor{black}{\iconTO} & \iconAppendOnly & \iconCOMMUNICATION{1}{0}{1}{0}{0}{1}{0} & \no &  & \iconCOMMUNICATION{1}{0}{0}{0}{0}{0}{0}[dotted, thick] & \yes & \makecell{sketch} & \redno & \redno & \iconCENTRALIZED[0] ~ \hspace*{-5ex} ~ \iconOperator & \iconCENTRALIZED[0] ~ \hspace*{-5ex} ~ \iconOperator &  & \makecell{Transfer\\ Digital Coin}
            \\ \hline 
1993 & \cite{Bra93} (TW E-Cash) & \iconNegativeInfo & \textcolor{black}{\iconTO} & \iconAppendOnly & \iconCOMMUNICATION{1}{0}{1}{0}{0}{0}{0} & \no &  & \iconCOMMUNICATION{1}{0}{0}{0}{0}{0}{0}[dotted, thick] & \yes & \makecell{sketch} & \redno & \redhalf & \iconCENTRALIZED[0] ~ \hspace*{-5ex} ~ \iconOperator & \iconCENTRALIZED[0] ~ \hspace*{-5ex} ~ \iconOperator &  & \makecell{Transfer\\ Digital Coin}
            \\ \hline 
1998 & \cite{CFT98} & \iconNegativeInfo & \textcolor{black}{\iconTO} & \iconAppendOnly & \iconCOMMUNICATION{1}{0}{1}{0}{0}{0}{0} & \no &  & \iconCOMMUNICATION{1}{0}{0}{0}{0}{0}{0}[dotted, thick] & \yes & \makecell{sketch} & \redno & \redno & \iconCENTRALIZED[0] ~ \hspace*{-5ex} ~ \iconOperator & \iconCENTRALIZED[0] ~ \hspace*{-5ex} ~ \iconOperator &  & \makecell{Transfer\\ Digital Coin}
            \\ \hline 
1999 & \cite{ST99a} (Flow Control) & \iconNegativeInfo & \textcolor{black}{\iconTO} & \iconAppendOnly & \iconCOMMUNICATION{0}{1}{0}{0}{0}{0}{0} & \no &  & \iconCOMMUNICATION{0}{1}{1}{0}{0}{0}{0}[dotted, thick] & \yes & \makecell{sketch} & \redno & \redno & \iconCENTRALIZED[0] ~ \hspace*{-5ex} ~ \iconOperator & \iconCENTRALIZED[0] ~ \hspace*{-5ex} ~ \iconOperator &  & \makecell{Transfer\\ Digital Coin}
            \\ \hline 
1999 & \cite{ST99b} (Auditable E-Cash) & \iconBothInfo & \textcolor{black}{\iconTO} & \iconPublicAvailable & \iconCOMMUNICATION{1}{0}{1}{0}{0}{0}{0} & \no &  & \iconCOMMUNICATION{1}{0}{0}{0}{0}{0}{0}[dotted, thick] & \yes & \makecell{sketch} & \redno & \redno & \iconCENTRALIZED[0] ~ \hspace*{-5ex} ~ \iconOperator & \iconCENTRALIZED[0] ~ \hspace*{-5ex} ~ \iconPublic &  & \makecell{Transfer\\ Digital Coin}
            \\ \hline 
2005 & \cite{CHL05} & \iconBothInfo & \textcolor{black}{\iconTO} & \iconAppendOnly & \iconCOMMUNICATION{1}{0}{1}{0}{0}{0}{0} & \no &  & \iconCOMMUNICATION{1}{0}{0}{0}{0}{0}{0}[dotted, thick] & \yes & \makecell{game} & \redno & \redno & \iconCENTRALIZED[0] ~ \hspace*{-5ex} ~ \iconOperator & \iconCENTRALIZED[0] ~ \hspace*{-5ex} ~ \iconOperator &  & \makecell{Transfer\\ Digital Coin}
            \\ \hline 
2008 & \cite{CG08} (Transferable E-Cash) & \iconNegativeInfo & \textcolor{black}{\iconTO} & \iconAppendOnly & \iconCOMMUNICATION{1}{0}{1}{0}{0}{1}{0} & \no &  & \iconCOMMUNICATION{1}{0}{0}{0}{0}{0}{0}[dotted, thick] & \yes & \makecell{game} & \redno & \redno & \iconCENTRALIZED[0] ~ \hspace*{-5ex} ~ \iconOperator & \iconCENTRALIZED[0] ~ \hspace*{-5ex} ~ \iconOperator &  & \makecell{Transfer\\ Digital Coin}
            \\ \hline 
2008 & \cite{Nak08} (Bitcoin) & \iconPositiveInfo & \textcolor{black}{\iconTO} & \iconPublicAvailable & \iconCOMMUNICATION{0}{1}{0}{1}{1}{0}{0} & \no &  & \iconCOMMUNICATION{0}{1}{1}{0}{0}{0}{0}[dotted, thick] & \no & \makecell{deployed} & \redno & \redno & \iconDECENTRALIZED[2] ~ \hspace*{-5ex} ~ \iconPublic & \iconDECENTRALIZED[2] ~ \hspace*{-5ex} ~ \iconPublic &  & \makecell{Burn and\\ Create}
            \\ \hline 
2009 & \cite{FPV09} & \iconNegativeInfo & \textcolor{black}{\iconTO} & \iconAppendOnly & \iconCOMMUNICATION{1}{0}{1}{0}{0}{1}{0} & \no &  & \iconCOMMUNICATION{1}{0}{0}{0}{0}{0}{0}[dotted, thick] & \yes & \makecell{game} & \redno & \redno & \iconCENTRALIZED[0] ~ \hspace*{-5ex} ~ \iconOperator & \iconCENTRALIZED[0] ~ \hspace*{-5ex} ~ \iconOperator & \iconCENTRALIZED[0] ~ \hspace*{-5ex} ~ \iconUser & \makecell{Transfer\\ Digital Coin}
            \\ \hline 
2010 & \cite{CG10} & \iconNegativeInfo & \textcolor{black}{\iconTO} & \iconAppendOnly & \iconCOMMUNICATION{1}{0}{1}{0}{0}{0}{0} & \no &  & \iconCOMMUNICATION{1}{0}{0}{0}{0}{0}{0}[dotted, thick] & \yes & \makecell{game} & \redno & \redno & \iconCENTRALIZED[0] ~ \hspace*{-5ex} ~ \iconOperator & \iconCENTRALIZED[0] ~ \hspace*{-5ex} ~ \iconOperator &  & \makecell{Transfer\\ Digital Coin}
            \\ \hline 
2013 & \cite{Sab13} (Monero) & \iconBothInfo & \textcolor{black}{\iconTO} & \iconPublicAvailable & \iconCOMMUNICATION{0}{1}{0}{1}{1}{0}{0} & \no &  & \iconCOMMUNICATION{0}{1}{1}{0}{0}{0}{0}[dotted, thick] & \yes & \makecell{game \\ deployed} & \redno & \redno & \iconDECENTRALIZED[2] ~ \hspace*{-5ex} ~ \iconPublic & \iconDECENTRALIZED[2] ~ \hspace*{-5ex} ~ \iconPublic &  & \makecell{Burn and\\ Create}
            \\ \hline 
2014 & \cite{BCG14} (Zerocash) & \iconBothInfo & \textcolor{black}{\iconTO} & \iconPublicAvailable & \iconCOMMUNICATION{0}{1}{0}{1}{1}{0}{0} & \no &  & \iconCOMMUNICATION{0}{1}{1}{0}{0}{0}{0}[dotted, thick] & \yes & \makecell{game \\ deployed} & \redyes & \redno & \iconDECENTRALIZED[2] ~ \hspace*{-5ex} ~ \iconPublic & \iconDECENTRALIZED[2] &  & \makecell{Burn and\\ Create}
            \\ \hline 
2014 & \cite{CPS14} (Divisible E-Cash) & \iconNegativeInfo & \textcolor{black}{\iconTO} & \iconAppendOnly & \iconCOMMUNICATION{1}{0}{1}{0}{0}{0}{0} & \no &  & \iconCOMMUNICATION{1}{0}{0}{0}{0}{0}{0}[dotted, thick] & \yes & \makecell{game} & \redyes & \redno & \iconCENTRALIZED[0] ~ \hspace*{-5ex} ~ \iconOperator & \iconCENTRALIZED[0] ~ \hspace*{-5ex} ~ \iconOperator &  & \makecell{Transfer\\ Digital Coin}
            \\ \hline 
2014 & \cite{But14} (Ethereum) & \iconPositiveInfo & \textcolor{black}{\iconTO} & \iconPublicAvailable & \iconCOMMUNICATION{0}{1}{0}{1}{1}{0}{0} & \no &  & \iconCOMMUNICATION{0}{1}{1}{0}{0}{0}{0}[dotted, thick] & \yes & \makecell{deployed} & \redno & \redno & \iconDECENTRALIZED[2] ~ \hspace*{-5ex} ~ \iconPublic & \iconDECENTRALIZED[2] ~ \hspace*{-5ex} ~ \iconPublic &  & \makecell{Global State\\ Update}
            \\ \hline 
2015 & \cite{BCF15} & \iconNegativeInfo & \textcolor{black}{\iconTO} & \iconAppendOnly & \iconCOMMUNICATION{1}{0}{1}{0}{0}{1}{0} & \no &  & \iconCOMMUNICATION{1}{0}{0}{0}{0}{0}{0}[dotted, thick] & \yes & \makecell{game} & \redyes & \redno & \iconCENTRALIZED[0] ~ \hspace*{-5ex} ~ \iconOperator & \iconCENTRALIZED[0] ~ \hspace*{-5ex} ~ \iconOperator &  & \makecell{Transfer\\ Digital Coin}
            \\ \hline 
2016 & \cite{GGM16} & \iconBothInfo & \textcolor{black}{\iconTO} & \iconPublicAvailable & \iconCOMMUNICATION{0}{1}{0}{1}{1}{0}{0} & \no &  & \iconCOMMUNICATION{0}{1}{1}{0}{0}{0}{0}[dotted, thick] & \yes & \makecell{game} & \redno & \redno & \iconDECENTRALIZED[2] ~ \hspace*{-5ex} ~ \iconPublic & \iconDECENTRALIZED[2] ~ \hspace*{-5ex} ~ \iconPublic & \iconCENTRALIZED[0] & \makecell{Burn and\\ Create}
            \\ \hline 
2016 & \cite{DM16} (RSCoin) & \iconPositiveInfo & \textcolor{black}{\iconTO} & \iconPublicAvailable & \iconCOMMUNICATION{0}{1}{0}{1}{1}{0}{1} & \half & \iconSync & \iconCOMMUNICATION{0}{1}{1}{0}{0}{0}{0}[dotted, thick] & \yes & \makecell{sketch \\ PoC} & \redno & \redno & \iconCENTRALIZED[0] ~ \hspace*{-5ex} ~ \iconPublic & \iconCENTRALIZED[1] ~ \hspace*{-5ex} ~ \iconPublic &  & \makecell{Burn and\\ Create}
            \\ \hline 
2019 & \cite{WKC19} (PRCash) & \iconPositiveInfo & \textcolor{black}{\iconTO} & \iconPublicAvailable & \iconCOMMUNICATION{1}{1}{0}{1}{1}{0}{0} & \half &  & \iconCOMMUNICATION{1}{1}{1}{0}{0}{0}{0}[dotted, thick] & \yes & \makecell{sketch \\ PoC} & \redno & \redno & \iconFEDERATED[2] ~ \hspace*{-5ex} ~ \iconPublic & \iconCENTRALIZED[0] ~ \hspace*{-5ex} ~ \iconPublic &  & \makecell{Burn and\\ Create}
            \\ \hline 
2019 & \cite{ACD19} & \iconBothInfo & \textcolor{black}{\iconTO} & \iconPublicAvailable & \iconCOMMUNICATION{0}{1}{0}{1}{0}{0}{0} & \half &  & \iconCOMMUNICATION{0}{1}{1}{0}{0}{0}{0}[dotted, thick] & \yes & \makecell{UC \\ PoC} & \redno & \redno & \iconFEDERATED[2] ~ \hspace*{-5ex} ~ \iconPublic & \iconFEDERATED[0] ~ \hspace*{-5ex} ~ \iconPublic & \iconCENTRALIZED[0] & \makecell{Burn and\\ Create}
            \\ \hline 
2020 & \cite{BDS20} (FastPay) & \iconPositiveInfo & \textcolor{black}{\iconNTO} & \iconPrivate & \iconCOMMUNICATION{0}{1}{0}{1}{0}{0}{0} & \yes & \iconAsync & \iconCOMMUNICATION{0}{0}{0}{0}{0}{0}{0}[dotted, thick] & \yes & \makecell{sketch \\ PoC} & \redno & \redno & \iconFEDERATED[2] & \iconDECENTRALIZED[0] ~ \hspace*{-5ex} ~ \iconOperator &  & \makecell{Global State\\ Update}
            \\ \hline 
2020 & \cite{CMT20} (PGC) & \iconPositiveInfo & \textcolor{black}{\iconTO} & \iconPublicAvailable & \iconCOMMUNICATION{0}{1}{0}{1}{1}{0}{0} & \no &  & \iconCOMMUNICATION{0}{1}{1}{0}{0}{0}{0}[dotted, thick] & \yes & \makecell{game \\ PoC} & \redno & \redno & \iconDECENTRALIZED[0] ~ \hspace*{-5ex} ~ \iconPublic &  & \iconCENTRALIZED[0] & \makecell{Global State\\ Update}
            \\ \hline 
2021 & \cite{WKD21} (Platypus) & \iconNegativeInfo & \textcolor{black}{\iconTO} & \iconAppendOnly & \iconCOMMUNICATION{1}{0}{1}{0}{0}{0}{0} & \no &  & \iconCOMMUNICATION{1}{1}{1}{0}{0}{0}{0}[dotted, thick] & \yes & \makecell{game \\ PoC} & \redyes & \redno & \iconCENTRALIZED[0] ~ \hspace*{-5ex} ~ \iconPublic & \iconCENTRALIZED[0] ~ \hspace*{-5ex} ~ \iconOperator &  & \makecell{Local State\\ Update}
            \\ \hline 
2021 & \cite{CGM21} (EC1) & \iconNegativeInfo & \textcolor{black}{\iconTO} & \iconAppendOnly & \iconCOMMUNICATION{1}{0}{1}{0}{0}{0}{1} & \no &  & \iconCOMMUNICATION{1}{0}{0}{0}{0}{0}{0}[dotted, thick] & \no & \makecell{PoC} & \redno & \redno & \iconCENTRALIZED[0] ~ \hspace*{-5ex} ~ \iconOperator & \iconCENTRALIZED[0] ~ \hspace*{-5ex} ~ \iconOperator &  & \makecell{Transfer\\ Digital Coin}
            \\ \hline 
2021 & \cite{BFQ21} & \iconNegativeInfo & \textcolor{black}{\iconTO} & \iconAppendOnly & \iconCOMMUNICATION{1}{0}{1}{0}{0}{1}{0} & \no &  & \iconCOMMUNICATION{1}{0}{0}{0}{0}{0}{0}[dotted, thick] & \yes & \makecell{game} & \redyes & \redno & \iconCENTRALIZED[0] ~ \hspace*{-5ex} ~ \iconOperator & \iconCENTRALIZED[0] ~ \hspace*{-5ex} ~ \iconOperator &  & \makecell{Transfer\\ Digital Coin}
            \\ \hline 
2022 & \cite{CM22} (EC2) & \iconPositiveInfo & \textcolor{black}{\iconTO} & \iconPublicAvailable & \iconCOMMUNICATION{1}{0}{1}{0}{0}{0}{1} & \no &  & \iconCOMMUNICATION{1}{1}{1}{0}{0}{0}{0}[dotted, thick] & \no & \makecell{PoC} &  & \redno & \iconCENTRALIZED[0] ~ \hspace*{-5ex} ~ \iconPublic & \iconCENTRALIZED[0] ~ \hspace*{-5ex} ~ \iconPublic &  & \makecell{\vphantom{Ag} \\ \vphantom{Ag}}
            \\ \hline 
2022 & \cite{TBA22} (Utt) & \iconNegativeInfo & \textcolor{black}{\iconNTO} & \iconPublicAvailable & \iconCOMMUNICATION{0}{1}{0}{1}{0}{0}{0} & \yes & \iconPartiallySync & \iconCOMMUNICATION{1}{1}{1}{0}{0}{0}{0}[dotted, thick] & \yes & \makecell{UC \\ PoC} & \redyes & \redno & \iconFEDERATED[2] & \iconFEDERATED[2] ~ \hspace*{-5ex} ~ \iconDedicatedEntity &  & \makecell{Burn and\\ Create}
            \\ \hline 
2022 & \cite{KKS22} (PEReDi) & \iconNegativeInfo & \textcolor{black}{\iconOccTO} & \iconAppendOnly & \iconCOMMUNICATION{1}{1}{1}{1}{1}{0}{0} & \yes & \iconAsync & \iconCOMMUNICATION{1}{1}{1}{0}{0}{0}{0}[dotted, thick] & \yes & \makecell{UC} & \redyes & \redno & \iconFEDERATED[2] & \iconCENTRALIZED[0] & \iconFEDERATED[2] & \makecell{Local State\\ Update}
            \\ \hline 
2022 & \cite{BSK22} (Zef) & \iconBothInfo & \textcolor{black}{\iconNTO} & \iconPrivate & \iconCOMMUNICATION{1}{1}{0}{1}{0}{0}{0} & \yes & \iconAsync & \iconCOMMUNICATION{1}{1}{1}{0}{0}{0}{0}[dotted, thick] & \no & \makecell{PoC} & \redno & \redno & \iconFEDERATED[2] & \iconDECENTRALIZED[0] ~ \hspace*{-5ex} ~ \iconOperator &  & \makecell{\vphantom{Ag} \\ \vphantom{Ag}}
            \\ \hline 
2023 & \cite{SKK23} (Parscoin) & \iconNegativeInfo & \textcolor{black}{\iconNTO} & \iconAppendOnly & \iconCOMMUNICATION{1}{1}{1}{1}{0}{0}{0} & \half &  & \iconCOMMUNICATION{0}{1}{1}{0}{0}{0}{0}[dotted, thick] & \yes & \makecell{UC} & \redyes & \redno & \iconFEDERATED[2] & \iconFEDERATED[2] & \iconFEDERATED[2] & \makecell{Local State\\ Update}
            \\ \hline 
2023 & \cite{RP23} & \iconNegativeInfo & \textcolor{black}{\iconTO} & \iconAppendOnly & \iconCOMMUNICATION{1}{0}{1}{0}{0}{1}{0} & \no & \iconAsync & \iconCOMMUNICATION{1}{0}{0}{0}{0}{0}{0}[dotted, thick] & \yes & \makecell{UC \\ PoC} & \redyes & \redno & \iconFEDERATED[2] ~ \hspace*{-5ex} ~ \iconOperator & \iconFEDERATED[2] ~ \hspace*{-5ex} ~ \iconOperator &  & \makecell{Transfer\\ Digital Coin}
            \\ \hline 
2023 & \cite{LVF23} (Hamilton) & \iconPositiveInfo & \textcolor{black}{\iconTO} & \iconPrivate & \iconCOMMUNICATION{1}{1}{0}{1}{0}{0}{1} & \half &  & \iconCOMMUNICATION{0}{1}{1}{0}{0}{0}{0}[dotted, thick] & \yes & \makecell{sketch \\ PoC} & \redno & \redno & \iconCENTRALIZED[1] ~ \hspace*{-5ex} ~ \iconOperator & \iconCENTRALIZED[0] ~ \hspace*{-5ex} ~ \iconOperator &  & \makecell{Burn and\\ Create}
            \\ \hline 
2024 & \cite{BZK24} (PayOff) & \iconNegativeInfo & \textcolor{black}{\iconNTO} & \iconPublicAvailable & \iconCOMMUNICATION{1}{0}{0}{0}{0}{1}{0} & \no &  & \iconCOMMUNICATION{1}{1}{1}{0}{0}{0}{0}[dotted, thick] & \yes & \makecell{sketch \\ PoC} & \redyes & \redhalf & \iconCENTRALIZED[0] ~ \hspace*{-5ex} ~ \iconPublic & \iconCENTRALIZED[0] & \iconCENTRALIZED[0] ~ \hspace*{-5ex} ~ \iconUser & \makecell{Local State\\ Update}
            \\ \hline 
2024 & \cite{ADE24} & \iconNegativeInfo & \textcolor{black}{\iconTO} & \iconAppendOnly & \iconCOMMUNICATION{1}{0}{1}{0}{0}{1}{1} & \no &  & \iconCOMMUNICATION{1}{0}{0}{0}{0}{0}{0}[dotted, thick] & \yes & \makecell{game \\ PoC} & \redno & \redhalf & \iconCENTRALIZED[0] & \iconCENTRALIZED[0] & \iconFEDERATED[0] & \makecell{Transfer\\ Digital Coin}
            \\ \hline 
2025 & \cite{BLK25} (PaxPay) & \iconBothInfo & \textcolor{black}{\iconNTO} & \iconAppendOnly & \iconCOMMUNICATION{1}{1}{0}{1}{0}{0}{0} & \half & \iconAsync & \iconCOMMUNICATION{1}{1}{1}{0}{0}{0}{0}[dotted, thick] & \yes & \makecell{game \\ PoC} & \redyes & \redno & \iconFEDERATED[2] & \iconFEDERATED[2] ~ \hspace*{-5ex} ~ \iconDedicatedEntity & \iconDECENTRALIZED[1] ~ \hspace*{-5ex} ~ \iconUser & \makecell{Burn and\\ Create}
            \\ \hline 
\end{tabular}
}
\label{tbl:main1}
\end{table*}

\renewcommand{\arraystretch}{1,95} 
\begin{table*}[!htpb]  
  \centering
  \caption[caption]{
    Part 2 of the classification of \numpapers{} designs according to Section~\ref{sec:categories} sorted by year of publication. Undesirable properties are shown in red.
  }
  \resizebox{\linewidth}{!}{
  \begin{tabular}{l|l||c|c|c|c||c|c||c|c|c|c|c|c||c|c|c}

\multicolumn{2}{c||}{\textbf{General info}} &
\multicolumn{4}{c||}{\textbf{Privacy}} &
\multicolumn{2}{c||}{\textbf{Compliance}} &
\multicolumn{6}{c||}{\textbf{Performance}} &
\multicolumn{3}{c}{\textbf{Technical features}} 
\\

Year &
Reference &
\makecell{Identity \\ privacy} &
\makecell{Value \\ privacy} &
\makecell{User \\ unlinkab.} &
\makecell{Sender \\ anon.} &
\makecell{Privacy \\ revoc.} &
\makecell{Limits} &
\makecell{Impl. \\ available} &
\makecell{Tx/s} &
\makecell{Latency} &
\makecell{Type of \\ benchmarks} &
\makecell{Sequential \\ payments} &
\makecell{Growing global \\ information} &
\makecell{Offline \\ payment} &
\makecell{Exact \\ amount} &
\makecell{Uninvolved \\ recipient}
\\ \hline

1982 & \cite{Cha82} (Blind Sig.) & \SP{} & \no & \no & \yes & \yes &  & \yes &   &  &  & \redno & \redyes & \no & \no & \no
           \\ \hline 
1988 & \cite{CFN88} (E-Cash) & \SP{} & \no & \no & \yes & \doublespender &  & \yes &   &  &  & \redno & \redyes & \half & \half & \no
           \\ \hline 
1991 & \cite{OO91} (Off-line E-Cash) & \SPo{} & \no & \no & \no & \doublespender &  & \no &   &  &  & \redno & \redyes & \yes & \half & \no
           \\ \hline 
1993 & \cite{Bra93} (TW E-Cash) & \SP{} & \no & \no & \yes & \doublespender &  & \no &   &  &  & \redno & \redyes & \half & \no & \no
           \\ \hline 
1998 & \cite{CFT98} & \SP{} & \no & \no & \no & \doublespender &  & \no &   &  &  & \redno & \redyes & \half & \half & \no
           \\ \hline 
1999 & \cite{ST99a} (Flow Control) & \SP{} & \no & \no & \yes & \no & S* & \no &   &  &  & \redno & \redyes & \no & \no & \yes
           \\ \hline 
1999 & \cite{ST99b} (Auditable E-Cash) & \SP{} & \no & \no & \yes & \doublespender &  & \no &   &  &  & \redrhalf & \redyes & \half & \no & \no
           \\ \hline 
2005 & \cite{CHL05} & \SP{} & \no & \no & \yes & \doublespender &  & \no &   &  &  & \redno & \redyes & \half & \no & \no
           \\ \hline 
2008 & \cite{CG08} (Transferable E-Cash) & \SPo{} & \no & \no & \yes & \doublespender &  & \no &   &  &  & \redno & \redyes & \yes & \no & \no
           \\ \hline 
2008 & \cite{Nak08} (Bitcoin) & \yes{} & \no & \no & \no & \no &  & \yes \iOSS & < 1000 & min & \iconLive & \redno & \redyes & \no & \yes & \yes
           \\ \hline 
2009 & \cite{FPV09} & \SPo{} & \no & \no & \yes & \unwanted &  & \no &   &  &  & \redno & \redyes & \yes & \no & \no
           \\ \hline 
2010 & \cite{CG10} & \SP{} & \no & \no & \yes & \doublespender &  & \no &   &  &  & \redno & \redyes & \half & \half & \no
           \\ \hline 
2013 & \cite{Sab13} (Monero) & \yes{} & \no & \half & \half & \no &  & \yes \iOSS & < 1000 & min & \iconLive & \redno & \redyes & \no & \yes & \yes
           \\ \hline 
2014 & \cite{BCG14} (Zerocash) & \yes{} & \yes & \yes & \yes & \no &  & \yes \iOSS & < 1000 & min & \iconLive & \redno & \redyes & \no & \yes & \yes
           \\ \hline 
2014 & \cite{CPS14} (Divisible E-Cash) & \SP{} & \no & \no & \yes & \doublespender &  & \no &   &  &  & \redno & \redyes & \half & \half & \no
           \\ \hline 
2014 & \cite{But14} (Ethereum) & \yes{} & \no & \no & \no & \no &  & \yes \iOSS & < 1000 & sec & \iconLive & \redno & \redyes & \no & \yes & \yes
           \\ \hline 
2015 & \cite{BCF15} & \SPo{} & \no & \no & \yes & \doublespender &  & \no &   &  &  & \redno & \redyes & \yes & \no & \no
           \\ \hline 
2016 & \cite{GGM16} & \yes{} & \yes & \yes & \yes & \yes & T, S & \no &   &  &  & \redno & \redyes & \no & \yes & \yes
           \\ \hline 
2016 & \cite{DM16} (RSCoin) & \yes{} & \no & \no & \no & \no &  & \yes \iOSS & < 1000 & ms & \iconSynthetic & \redno & \redyes & \no & \yes & \yes
           \\ \hline 
2019 & \cite{WKC19} (PRCash) & \yes{} & \half & \no & \no & \no & S, R & \partially & > 1000 & ms & \iconSynthetic & \redno & \redyes & \no & \yes & \no
           \\ \hline 
2019 & \cite{ACD19} & \yes{} & \yes & \yes & \yes & \yes &  & \partially &   &  &  & \redno & \redyes & \no & \yes & \yes
           \\ \hline 
2020 & \cite{BDS20} (FastPay) & \no{} & \no & \no & \no & \no &  & \yes \iOSS & > 100 000 & ms & \iconSynthetic & \redhalf & \redno & \no & \yes & \yes
           \\ \hline 
2020 & \cite{CMT20} (PGC) & \yes{} & \half & \no & \no & \yes & S, R & \partially \iOSS &   &  &  & \redhalf & \redyes & \no & \yes & \yes
           \\ \hline 
2021 & \cite{WKD21} (Platypus) & \yes{} & \yes & \yes & \yes & \no & H, S, R & \partially & < 1000 &  & \iconSynthetic & \redyes & \redyes & \no & \yes & \no
           \\ \hline 
2021 & \cite{CGM21} (EC1) & \SP{} & \no & \no & \yes & \no &  & \yes & > 1000 & sec & \iconSynthetic & \redno & \redyes & \no & \half & \no
           \\ \hline 
2021 & \cite{BFQ21} & \SPo{} & \no & \no & \yes & \doublespender &  & \no &   &  &  & \redno & \redyes & \yes & \half & \no
           \\ \hline 
2022 & \cite{CM22} (EC2) & \SP{} & \no & \no & \yes & \no &  & \yes & > 1000 & sec & \iconSynthetic & \redno & \redno & \no & \no & \no
           \\ \hline 
2022 & \cite{TBA22} (Utt) & \yes{} & \yes & \yes & \yes & \no & S* & \yes \iOSS & > 1000 & ms & \iconSynthetic & \redno & \redyes & \no & \yes & \yes
           \\ \hline 
2022 & \cite{KKS22} (PEReDi) & \yes{} & \yes & \yes & \yes & \yes & S, R, H, T & \no &   &  &  & \redyes & \redyes & \no & \yes & \no
           \\ \hline 
2022 & \cite{BSK22} (Zef) & \yes{} & \yes & \yes & \yes & \no &  & \yes \iOSS & < 1000 & ms & \iconSynthetic & \redno & \redyes & \no & \yes & \yes
           \\ \hline 
2023 & \cite{SKK23} (Parscoin) & \yes{} & \yes & \yes & \yes & \yes &  & \no &   &  &  & \redyes & \redyes & \no & \yes & \no
           \\ \hline 
2023 & \cite{RP23} & \SP{} & \no & \no & \yes & \doublespender &  & \partially \iOSS &   &  &  & \redno & \redyes & \no & \half & \no
           \\ \hline 
2023 & \cite{LVF23} (Hamilton) & \yes{} & \no & \no & \no & \no &  & \yes \iOSS & > 1 000 000 & ms & \iconSynthetic & \redno & \redno & \no & \yes & \yes
           \\ \hline 
2024 & \cite{BZK24} (PayOff) & \yes{} & \yes & \yes & \yes & \unwanted & H, R, S & \partially & > 100 000 &  & \iconSynthetic & \redno & \redyes & \yes & \yes & \no
           \\ \hline 
2024 & \cite{ADE24} & \SPo{} & \no & \no & \yes & \yes &  & \partially &   &  &  & \redno & \redyes & \yes & \half & \no
           \\ \hline 
2025 & \cite{BLK25} (PaxPay) & \yes{} & \yes & \yes & \yes & \no & S, T & \yes & < 1000 & ms & \iconSynthetic & \redno & \redyes & \no & \yes & \yes
           \\ \hline 
\end{tabular}
\label{tbl:main2}
} 
\end{table*}

\subsection{Design Patterns} 
\label{ssec:data_patterns}

Our analysis revealed that the design space of all the digital payment systems reviewed in this paper can be broken down into four design patterns: \emph{Global State Update}, \emph{Local State Update}, \emph{Transfer Digital Coin}, and \emph{Burn-and-Create}. 
These four patterns capture three fundamental choices in system design: how value is represented, how a payment is authorised, and how it is validated.

The most straightforward payment system design is a mapping where a current balance is assigned to each user. 
A payment is executed by updating the balances of the involved parties accordingly. 
We call designs based on this idea \emph{Global State Update}. 
Here, value is represented by a global state that specifies the distribution of value (i.\,e., positive information), and every payment updates this global state. 
Authorisation requires a secret, typically a private key or knowledge of credentials. 
Validation involves checking a signature or credentials, and confirming that the sender has sufficient funds according to the global state. 
If validation succeeds, the global state is updated. 
This approach corresponds to the familiar account-based model used in the current banking system and is also employed in cryptocurrencies, such as Ethereum~\cite{But14}. 

Another approach shifts balance tracking from the operator to the users, who each maintains their own account record.
We call this pattern \emph{Local State Update}. 
For each transaction, the users update their local accounts and generate zero-knowledge proofs showing that the update is correct, that the previous state was signed by the operator, and that they are authorised.
The new local states remain hidden. 
The users submit commitments to the operator, the proofs, and references to the previous state.
Validation involves verifying the proofs and ensuring that the referenced global state has not been reused, thereby preventing users from reverting to earlier states with higher balances.
If validation succeeds, the operator signs the new state commitment and records the old reference in their list of past states (i.\,e., negative information).
This approach was first proposed in Platypus~\cite{WKD21} and extended in subsequent work.
Earlier variants of this pattern that rely on trusted hardware rather than zero-knowledge proofs were used in prepaid card systems in the 1990s, such as Quick in Austria and Geldkarte in Germany.
However, these cards could only be used to spend funds at payment terminals, not to receive them from other cards~\cite{bis2000electronicMoney}.

When paying with cash, we simply transfer a piece of paper or metal. 
Chaum~\cite{Cha82} brought this concept into the digital world by proposing a system in which the payer sends a confidential bit string to the recipient. 
We call this pattern \emph{Transfer Digital Coin}.
Users obtain a signature on a blinded bit string from the bank, which assigns value to it. 
A payment consists of transferring this signed, confidential bit string to another party, who can then redeem it at the bank. 
The bank verifies the signature and checks that the bit string has not already been redeemed.
It then records the string in a list of redeemed coins (i.\,e., negative information). 

A different way of representing value was introduced in the Bitcoin whitepaper~\cite{Nak08}. 
An \emph{unspent transaction output} (UTXO) is a tuple of amount and spending condition, typically a signature matching the owner's public key.
Crucially, each party can own multiple UTXOs.
The global state records all currently valid UTXOs, i.\,e., positive information. 
With each payment, one or more UTXOs are removed from the global state, or \emph{burned}, to create new UTXOs of equal total amount, which are added to the global state. 
We call this pattern \emph{Burn-and-Create}. 
A payment is authorised by fulfilling the spending condition of the input UTXOs, while validation consists of checking the signature and ensuring that all input values existed prior to being burned. 
A variant of this pattern uses positive and negative information.
In this case, a payment does not burn inputs directly but rather commitments to the inputs. 
A zero-knowledge proof is required to show that the commitments open to valid values in the global state.
Validation then requires checking the zero-knowledge proof and the signature that authenticates spending, as well as confirming that the commitments have not already been spent. 
Once validated, the burned commitments are added to the list of used commitments (negative information), while the newly created values are added to the list of UTXOs (positive information).
This variant of \emph{Burn-and-Create} was partially implemented in Monero~\cite{Sab13} and later realised in Zerocash~\cite{BCG14}. 

Two designs that do not fully fit into either of the design patterns are Zef~\cite{BSK22} and the second phase of Project Tourbillion~\cite{CM22}.
Zef has a base system that follows the \emph{Global State Update}, and for anonymous payments, a system resembling \emph{Burn and Create}.
EC2~\cite{CM22} builds on the \emph{Transfer Digital Coin} pattern, but switches from negative to positive information by using mixes~\cite{DBLP:journals/cacm/Chaum81} instead of blinding the bitstring.
While this positively affects performance, the provided privacy depends solely on the integrity of the mixes.

\subsection{Quantitative Perspective}
\label{ssec:pattern_quant}
Table~\ref{tab:data_patterns} provides an overview of the relationship between the presented design patterns and privacy, technical features, and performance properties.
First, the table shows the number of analysed papers belonging to each category.
The table shows the percentage of papers for each design pattern that offer identity privacy, value privacy, user unlinkability and sender anonymity.
Two pies in the Identity Privacy column indicate that the identity privacy for the sender (left) and the recipient (right) differ.
The striped area indicates that identity privacy applies only in the case of an offline payment.
For value privacy, the striped area corresponds to papers that hide only values (see Section~\ref{sssec:value_privacy}).
For user unlinkability and sender anonymity, the striped areas indicate set unlinkability (see Section~\ref{sssec:user_unlinkability} and Section~\ref{sssec:sender_anon}).
Next, the table shows the percentage of papers within each pattern that achieve the technical features Offline Payment, Exact Amounts and Uninvolved Recipient.
For the offline category, striped areas mean one-time offline functionality (see Section~\ref{sssec:offline}).
For exact amounts, they indicate that payments of varying amounts, but not any amount, are possible (see Section~\ref{sssec:exact}).
Finally, the table shows, for each pattern, the number of papers featuring an infinitely growing global state and whether payments are restricted to be sequential.
Striped areas in the sequential payment category represent papers that impose constraints on either the sender or the recipient only.

The table shows that none of the designs is perfect across properties.
Essentially, \emph{Global State Update} designs lack privacy and offline functionality, \emph{Local State Update} designs still require efforts to improve performance, \emph{Transfer Digital Coin} designs lack privacy for the recipient and require measures to enable exact payments.
The \emph{Burn and Create} designs either lack privacy or performance, and both do not support offline functionality.

\begin{table}[t]
    \centering
    \caption{Overview of the design patterns and the share of corresponding designs that provide the different privacy, technical features and performance properties. Undesirable properties are marked in red.}
    \small
    \renewcommand{\arraystretch}{1.2}
        \begin{tabular}{ccccccccccc}
            \toprule
            \makecell{\textbf{Design}\\\textbf{pattern}} &
            \rotatebox{90}{\makecell[l]{\textbf{Number of}\\\textbf{designs}}} &
            \rotatebox{90}{\makecell[l]{\textbf{Identity}\\\textbf{privacy}}} &
            \rotatebox{90}{\makecell[l]{\textbf{Value}\\\textbf{privacy}}} &
            \rotatebox{90}{\makecell[l]{\textbf{User}\\\textbf{unlinkability}}} &
            \rotatebox{90}{\makecell[l]{\textbf{Sender}\\\textbf{anonymity}}} &
            \rotatebox{90}{\makecell[l]{\textbf{Offline}\\\textbf{payment}}} &
            \rotatebox{90}{\makecell[l]{\textbf{Exact}\\\textbf{amounts}}} &
            \rotatebox{90}{\makecell[l]{\textbf{Uninvolved}\\\textbf{recipient}}} &
            \rotatebox{90}{\makecell[l]{\textbf{Infinitely}\\\textbf{growing}}} &
            \rotatebox{90}{\makecell[l]{\textbf{Sequential}\\\textbf{payments}}} \\
            \midrule 
            \makecell{\emph{Global State}\\ \emph{Update}} & 3 & \share{67}{0} & \share{0}{33} & \share{0}{0} & \share{0}{0}  & \share{0}{0} & \share{100}{0} & \share{100}{0} & \share{66}{0}[1] & \share{0}{66}[1] \\ \hline
            
            \makecell{\emph{Local State}\\ \emph{Update}} & 4 & \share{100}{0} & \share{100}{0} & \share{100}{0} & \share{100}{0}  & \share{25}{0} & \share{100}{0} & \share{0}{0} & \share{100}{0}[1] & \share{75}{0}[1] \\ \hline
            
            \makecell{\emph{Transfer}\\ \emph{Digital Coin}} & 17 & \share{100}{0}\hspace*{-1.5ex}\share{0}{35} & \share{0}{0} & \share{0}{0} & \share{88}{0}  & \share{35}{41} & \share{0}{53} & \share{6}{0} & \share{100}{0}[1] & \share{0}{6}[1] \\ \hline
            
            \multirow[b]{2}{*}{\makecell{\emph{Burn and}\\ \emph{Create}$^{\mathrm{a}}$}}
            & 4 & \share{100}{0} & \share{0}{25} & \share{0}{0} & \share{0}{0} & \share{0}{0} & \share{100}{0} & \share{75}{0} & \share{75}{0}[1] & \share{0}{0}[1] \\
            
            & 6 & \share{100}{0} & \share{83}{0} & \share{83}{17} & \share{83}{17} & \share{0}{0} & \share{100}{0} & \share{100}{0} & \share{100}{0}[1] & \share{0}{0}[1] \\
            \bottomrule
            \multicolumn{11}{p{\dimexpr\linewidth-2\tabcolsep\relax}}{
                \scriptsize $^{\mathrm{a}}$ The two rows for \emph{Burn and Create} show the two different versions. The one above is the classic version, like Bitcoin, and the one below uses zk-proofs.
            }
        \end{tabular}
    \label{tab:data_patterns}
\end{table}
\newpage
\section{Discussion}
\label{sec:discussion}

We evaluate the privacy, technical features with usability-implications and performance resulting from the design patterns discussed in Section~\ref{ssec:data_patterns}.

\subsection{Privacy Enabled by the Design Patterns}
\label{ssec:privacy_res}
It is convenient to discuss privacy along the design patterns because the way values is represented and updated impacts what information is observable by the operator.

The \emph{Global State Update} designs generally cannot hide the transaction graph; therefore, they do not provide user unlinkability or sender anonymity. 
Designs in this category offer either no privacy~\cite{BDS20}, identity privacy~\cite{But14}, or identity privacy combined with value privacy~\cite{CMT20}. 

The \emph{Local State Update} designs, when implemented with zero-knowledge proofs, support all privacy properties of our framework, namely identity privacy, value privacy, and user unlinkability. 

The \emph{Transfer Digital Coin} designs generally provide identity privacy for the sender and sender anonymity. 
Exceptions are early works~\cite{OO91,CFT98}, which focused on offline functionality and divisibility, respectively, and did not provide sender anonymity. 
\emph{Transfer Digital Coin} designs that support offline functionality achieve a stronger level of privacy, since they also provide identity privacy for the recipient across all offline hops. 

Finally, the \emph{Burn-and-Create} pattern has two different privacy levels. 
The original design pattern, as proposed in Bitcoin, cannot hide the transaction graph and only provides identity privacy~\cite{Nak08,DM16,LVF23} or, in some cases, partial value privacy~\cite{WKC19}. 
The variant combining positive and negative information supports all privacy properties of our framework~\cite{BCG14,GGM16,ACD19,TBA22,BLK25}. 
An exception is Monero~\cite{Sab13}, which provides set unlinkability and, in its current form~\cite{Monero2024Docs}, also value privacy. 

\subsection{Barriers to High Performance}
\label{ssec:performance_short}

Although latency and throughput depend on the full set of implementation details, such as the consensus mechanism (see Appendix~\ref{xsec:consensus_and_performance}), we identify some characteristics of the design pattern that impact performance.

\emph{Global State Update} systems can be efficient in principle, in particular when privacy is limited and there is no latency by slow cryptography.
In some cases, the sender must make sequential payments.
As this pattern relies on positive global information, the global state does not need to grow indefinitely.
However, designs with decentralized operation have higher consensus overhead and retain the entire transaction history for trustless validation.
Hence, not all systems using this pattern have high performance.

\emph{Local State Update} systems require complex cryptography, such as zero-knowledge proofs, and rely on a constantly growing global state.
While in early designs sequential payments per user are necessary, Payoff~\cite{BLK25} shows that these constraints are not inherent to the pattern.

\emph{Transfer Digital Coin} systems can be built with efficient cryptography, but they require infinitely growing global information.
There are generally no sequential payments necessary, with the exception of~\cite{ST99b}, where a recipient can receive only one coin per epoch.

The two \emph{Burn-and-Create} variants differ in terms of performance.
The variant with positive information promises the best performance of all patterns since the UTXO model can process concurrent payments involving the same users in parallel.
The variant that relies on both positive and negative information uses more complex cryptography due to zero-knowledge proofs.
It also requires an infinitely growing global state. 
As both positive and negative information continue to grow, creating the zero-knowledge proof becomes increasingly expensive (but at a decreasing rate).
Neither sender nor recipient are restricted to make payments sequentially.

We believe that all four patterns can be implemented in designs that, in principle, can perform at the scale of a CBDC.
Additional performance mechanisms, such as sharding (see Appendix~\ref{xsec:sharding}), can facilitate this.
\emph{Local State Update} is the most recent design pattern and reliable performance measurements are missing.

\subsection{Barriers to Usability} 
\label{ssec:usability_short}

We also find a relationship between the design pattern and possible technical features (as defined in Section~\ref{ssec:tech_feat}).

The literature suggests that systems following the \emph{Global State Update} pattern cannot support offline payments.
On the upside, they allow paying exact amounts and enable transactions without involving the recipients.
PayOff~\cite{BLK25} is the first (and only) \emph{Local State Update} design with offline functionality.
While \emph{Local State Update} designs enable paying exact amounts, all published systems require that the recipient is involved in the transaction process.
Many \emph{Transfer Digital Coin} designs provide offline functionality.
However, paying exact amounts remains challenging (see Appendix~\ref{xsec:exact_amounts}).
In offline mode, the recipient must be involved. 
Sanders and Ta-Shma~\cite{ST99a} demonstrate that this is not necessary online.
The \emph{Burn-and-Create} pattern does not support offline payments, whereas paying exact amounts and keeping the recipients uninvolved is straightforward.

Since all patterns have shortcomings, a fully usable CBDC may require two technical backends, one to support offline functionality and a second to handle push payments without involvement of the recipient.
Both parts must allow payments of exact amounts.
The online system can be implemented using either the \emph{Burn and Create} or the \emph{Global State Update} pattern.
Potential candidates for the offline system are based on the \emph{Transfer Digital Coin} and \emph{Local State Update} designs.
\section{Conclusion}
\label{sec:conclusion}
This paper presents the first top-down systematisation of knowledge on payment system designs suitable for CBDCs at this level of technical depth.
Beyond analysing \numpapers{} designs, we defined categories to enable comparison and identified recurring patterns, challenges, and trade-offs in the current state of the art.
The main challenges in the design space are as follows.

Balancing privacy and compliance seems to be contentious~\cite{auer_PETS_2025}, but technical solutions can be found once the policy discussion has converged.
However, two open problems remain. 
First, no system offering more than identity and value privacy has yet demonstrated scalability in practice.
Second, the development and rollout of a reliable, privacy-preserving identity system---which is needed to enforce compliance---remains a challenge.
It requires a strategic decision whether central banks should do this on their own or leverage national digital identity systems.

Quantum computers exploit the effects of quantum physics to perform computations efficiently that are expensive for classical computers.
It remains unclear when this technology will reach practical application, for instance, to break cryptographic primitives that are currently considered secure. 
The German Federal Office for Information Security (BSI) assumes that this could happen in the early 2030s~\cite{BSI_quantum_safe_crypto}. 
Practical quantum computers could pose a threat to the privacy and, even more critically, the integrity of digital payment systems~\cite{auer2024quantumcomp}. 
Nevertheless, only three analysed designs 
explicitly mention post-quantum considerations.
Project Tourbillon~\cite{CM22} is the only design with a post-quantum secure fallback, but it performs poorly in this setting.
Research on post-quantum secure payment systems remains insufficiently explored and unresolved.
A CBDC should implement a fallback system to ensure integrity, even in the presence of quantum computers, and it should be possible to upgrade privacy to post-quantum secure primitives.

Another rather underexplored area is offline payments.
Some \emph{Transfer Digital Coin} designs implemented this functionality in the 1990s, alongside trusted-hardware-based systems~\cite{GeldKarte}.
The only design following a different pattern, implementing the offline functionality, is PayOff~\cite{BLK25}.
Most importantly, none of the offline designs analysed in this systematisation have been implemented.
More research in this field is needed to identify the most suitable design for a CBDC with offline functionality.

Finally, a major gap in the field is the lack of comparable performance measures.
Privacy-friendly payment systems with compliance measures have yet to be implemented, as have post-quantum secure designs and offline payments.
Most of these reviewed designs originate from academic research, adding new functionality to existing concepts or improving upon them.
Authors rarely implement and measure full systems, including all components required for practical deployment.
Communication mixes and anonymous network layers are often assumed, but rarely realised in performance benchmarks.
Some proposals depend on underlying architectures, such as blockchain technology or commercial banking systems.
Translating a protocol into a practical system is challenging when key assumptions, such as synchrony, reliability, or trust, remain implicit or unspecified.
Moreover, none of the designs we analysed employ machine-verifiable proofs, rendering their security guarantees unverifiable without implementation.
However, the challenge does not end when a system is implemented.
Many benchmarks provided by designs with performance data may not reflect real-world retail usage, such as seasonal peaks around public holidays.
Of the \numpapers{} analysed, only 15 provide performance measurements, including benchmarks and, in most cases, latency.
Four of the designs are deployed cryptocurrencies; the remaining ones are synthetic benchmarks tailored to the specific design.
To the best of our knowledge, there is no standardised transaction dataset available for benchmarking payment systems, such as retail CBDCs.
Such a dataset would be essential for making reliable and reproducible comparisons, especially when assessing the performance of privacy-focused designs.

CBDCs are a new form of digital money with broad societal and economic impact.
If there is one thing this SoK has shown then it is the complexity of the problem and the breadth of the design space.
In the absence of a universally accepted definition of a CBDC, it is impossible for this work to present the single perfect CBDC design.
What an appropriate CBDC looks like depends fundamentally on the policy objectives.
Many of the proposals analysed in this work might serve as a potential stepping stones towards future CBDC systems.
As seen with TLS, which evolved over decades to reach today's robustness, CBDCs will also require continuous refinement.
Progress will depend on iterative, transparent development informed by real-world deployment and feedback.
We believe that such an approach is essential for creating secure, resilient, and widely trusted digital currencies.

\section*{Acknowledgment}

J.\;S.\ and R.\,B.\  thank the Austrian Research Promotion Agency (FFG) for funding our research project ``DeFiTrace'' and the Anniversary Fund of the Oesterreichische Nationalbank for supporting the research project on ``Privacy and the Functions of Digital Money,'' project number 18613.
All authors thank Verena Lachner, Kristina Magnussen and Martin Summer for providing useful comments on earlier drafts of this paper.

\bibliographystyle{references/style}
\bibliography{references/bibliography, references/CBDC_table}

\appendix
\section{Method}
\label{xsec:method}

In this appendix, we describe our approach and the procedure used to gather relevant literature, screen it, and select the final set of designs for the systematisation.

\subsection{Approach}
The aim of this work is to complement the growing number of publications reviewing existing PETs as building blocks for future payment systems, with a systematisation of proposals for complete systems that leverage privacy-enhancing features.
This helps to form intuitions on how privacy features can be combined in systems, working around the often-overlooked problem that PETs are not always easily composable.
Reviewing complete systems also allows for a more comprehensive view of the full set of relevant properties, notably performance and compliance.

Our process involved a literature search and selection (Section~\ref{xssec:litsel}) and an iterative development of a suitable categorisation.
Section~\ref{sec:categories} documents the resulting final categorisation.
The final set of $\numpapers$ systems is mapped to the classification, leading to the overview Tables~\ref{tbl:main1} and \ref{tbl:main2} that are at the heart of every SoK.
In our opinion, not every system in this table is a ``CBDC candidate,'' but every system marks an important point in the design space.
All systems were included because making an informed decision about any potential CBDC design requires an understanding of the alternatives and the reasons why some of them are more or less suitable.

\subsection{Literature Selection}
\label{xssec:litsel}
In order to be included in our systematisation, proposals must specify a complete payment system and include sufficient technical details to enable us to understand the relevant steps of the underlying protocols.
The goal is to review research contributions, i.\,e., technical documents describing a novel approach supported with evidence.
Most works are peer-reviewed. 

For the initial literature search, we used the search terms \texttt{private, regulated, scalable CBDC design} and \texttt{private, regulated, scalable central bank digital currency design} in October 2024 via the bibliographic database Google Scholar.\footnote{Cf. \url{https://scholar.google.com/}}
We sorted results by relevance without applying any additional filters. 
From the union of the first 100 results of each query, we identified 21 papers that proposed a concrete design for a CBDC or digital payment system. 
These papers underwent an initial assessment.
We determined that seven of them met the criterion of offering sufficient technical details.

Next, we performed a one-step forward and backward search by examining both the references of the seven initially selected papers and the papers that cited them. 
We repeated this process until we reached saturation.

While our literature search started broad, we had to exclude many of the economics papers, which often explore the monetary policy implications of CBDC adoption and the role of commercial banks. 
While some of these works include high-level design concepts, they generally lack the technical depth required for the analysis conducted in this SoK.
The same was true for reports of CBDC pilot projects, research initiatives, and already launched CBDCs. 
Note, that for Project Tourbillon~\cite{project_tourbillion}, we reference \cite{CMT20} (EC1) and \cite{CM22} (EC2) for phase 1 and phase 2, respectively. 

After this selection process, we were left with a collection of 66 design proposals.
Subsequently, these proposals were manually reviewed and certain designs excluded based on the following rationales:
\begin{itemize}
    \item Papers that were shown to be insecure (e.\,g., Zerocoin~\cite{MGG13});
    \item Publications centred around the same design if others were already included (e.\,g., early papers by David Chaum~\cite{Cha85}, since similar versions by the same author~\cite{Cha82, Cha88} were already considered);
    \item Designs that were significantly improved in follow-up work (e.\,g.,~\cite{OO89} by Okamoto was replaced by its improved version~\cite{OO91}).
\end{itemize}

This process resulted in a final set of \numpapers{} design proposals, which we analysed in depth.

\section{Communication Symbols}
\label{xsec:communication}

\begin{figure}[ht]
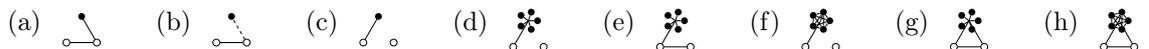
 
    \centering 
    (a)~\iconCOMMUNICATION{1}{0}{1}{0}{0}{0}{0} \hfill 
    (b)~\iconCOMMUNICATION{1}{0}{1}{0}{0}{1}{0} \hfill 
    (c)~\iconCOMMUNICATION{0}{1}{0}{0}{0}{0}{0} \hfill 
    (d)~\iconCOMMUNICATION{0}{1}{0}{1}{0}{0}{0} \hfill
    (e)~\iconCOMMUNICATION{1}{1}{0}{1}{0}{0}{0} \hfill
    (f)~\iconCOMMUNICATION{0}{1}{0}{1}{1}{0}{0} \hfill 
    (g)~\iconCOMMUNICATION{1}{1}{1}{1}{0}{0}{0} \hfill 
    (h)~\iconCOMMUNICATION{1}{1}{1}{1}{1}{0}{0} \\
    \caption{Different communication relations.} 
    \label{fig:communication-types} 
\end{figure}

In this appendix, we provide system examples corresponding to all communication patterns identified in our literature selection.
The complete set of patterns is depicted in Fig.~\ref{fig:communication-types}.
The bottom left node, always depicts the sender, the bottom right node the recipient and the top node the operator(s).  

Each communication pattern can be illustrated using existing systems.
Fig.~\ref{fig:communication-types}\,(a) shows the communication necessary in Chaumian e-cash (see, for example, \cite{Cha82}), where the sender transfers a coin to the recipient, who then deposits it with the operator, i.\,e., the bank.

Fig.~\ref{fig:communication-types}\,(b) shows transferable e-cash (e.\,g., \cite{OO91}), where the sender transfers a coin to the recipient, who can then transfer it to another user until it is deposited with the operator.

Fig.~\ref{fig:communication-types}\,(c) depicts Flow Control~\cite{ST99a}, where the sender sends a coin to the bank via an anonymous channel, including an identifier of the recipient, allowing the bank to credit the recipient's account.

Fig.~\ref{fig:communication-types}\,(d) shows FastPay~\cite{BDS20}, where the sender broadcasts a payment request to multiple operators.
Each operator individually signs the request, and once the sender collects enough signatures, these are broadcast again.
The operators then settle the payment locally, therefore not needing to communicate with each other.  

Fig.~\ref{fig:communication-types}\,(e) illustrates Zef~\cite{BSK22}, where the recipient first communicates the desired coins to the sender.
The sender then requests these coins from the operators, who each sign the request.
The sender aggregates the signatures and returns the signed coins to the recipient.

Fig.~\ref{fig:communication-types}\,(f) represents Bitcoin, where a user broadcasts a transaction request.
The operators, i.\,e., mining nodes, create a new block containing the transaction and broadcast it to all other nodes.

Fig.~\ref{fig:communication-types}\,(g) corresponds to PARScoin~\cite{SKK23}.
As in FastPay, operators sign state updates individually and do not communicate with each other.
However, unlike FastPay, the sender and recipient must communicate directly to update their respective states and then interact individually with the operators to sign the new states.

Finally, Fig.~\ref{fig:communication-types}\,(h) shows PEReDi~\cite{KKS22}, which is similar to PARScoin but requires operator-to-operator communication in the event of a transaction abort.

\section{Efficient Storage of Negative Information} 
\label{xsec:bloom}
This appendix discusses the engineering challenges of implementing nullifier-based approaches, i.\,e. designs having ever-growing lists of negative global information, and provides some estimates regarding the amount of data to be handled to meet desirable performance goals. 

Bloom filters are a probabilistic data structure which can be used to test whether a given element is a member of a known set in constant-time~\cite{DBLP:journals/cacm/Bloom70}. 
Thereby, they never provide \emph{false-negatives}, i.\,e., if an element is identified as not being part of the set it definitely is not. 
On the other hand, \emph{false-positives} are possible, i.\,e., if an element is identified as being part of the set there is still a probability that it is not. 
The rate of false positives is parameterised, and this parametrisation leads to trade-offs regarding size and the efficiency of the constant time query. 
Over the years, multiple different variants and optimisation techniques have been developed (see~\cite{LGMRL19_bloom} for a survey). 
To provide estimates regarding filter sizes and the required number of hash functions (or invocations) influencing the query time, we only consider a basic bloom filter design in this section. 
Table~\ref{tab:bloom} provides some calculations to estimate the size requirements of a bloom filter for fast non-membership proofs required to prevent double-spending. 
The table includes performance goals that would satisfy the current global average demand in the number of transactions and a maximum holding time of one year, i.\,e., only the total number of transactions in a year is considered.
Our estimates show that the number of hash function invocations, which is around $30$, is insignificant given that SHA3~\cite{NISTIR7896}, for example, can run in around 10 cycles per byte~\cite{bernstein2012sha3opt}. 
The required size of the bloom filter of around 19 TB is substantial given that it has to be held in memory for fast access. 
This size is not beyond reach for high-end infrastructures, but splitting up the filter into several parts has to be considered to distribute the load. 
For our calculations, we specified a false-positive rate of approximately one false-positive per day, which we consider acceptable. 
In any case, the use of bloom filter implies that there also has to be a mechanism to rule out false-positives using a slow-path. 
\begin{table}[ht]
\centering
\caption{Table of different bloom filter parametrisation to estimate the required size of the filter in a year for a given number of transactions per second and a false positive rate of approx. one false positive per day.}
\small
\begin{tabular}{rlrrr}
\toprule
Tx/sec. & Entries/year & False positive rate & Size & Hash functions \\
\midrule
20 & 6.31e+08 (630.7M) & 5.79e-07 (1 in 1,728,000) & 2.19 GB & 21 \\
100 & 3.15e+09 (3.2B) & 1.16e-07 (1 in 8,640,000) & 12.20 GB & 23 \\
1,000 & 3.15e+10 (31.5B) & 1.16e-08 (1 in 86,400,000) & 139.64 GB & 26 \\
50,000 & 1.58e+12 (1.6T) & 2.31e-10 (1 in 4,320,000,000) & 8.28 TB & 32 \\
110,000 & 3.47e+12 (3.5T) & 1.05e-10 (1 in 9,504,000,000) & 18.86 TB & 33 \\
1,000,000 & 3.15e+13 (31.5T) & 1.16e-11 (1 in 86,400,000,000) & 187.91 TB & 36 \\

\bottomrule
\end{tabular}
\label{tab:bloom}
\end{table}

\section{Consensus Performance Considerations}
\label{xsec:consensus_and_performance}

Agreement protocols such as consensus or state machine replication (SMR)\footnote{The exact delineation and definition of different agreement problems is complex and nuanced. We refer the reader to Garay and Kiayias\cite{garay2020sok} for a systematisation on the topic.} are an essential component of many of the distributed system designs of CBDCs that are considered in this work, providing strong consistency and fault-tolerance guarantees.
However, such agreement protocols are often associated with non-trivial communication and coordination costs, which can hinder throughput and scalability~\cite{guerraoui2019consensus}.

For example, Byzantine fault tolerance in SMR was long regarded as too inefficient for practical deployment, until Castro and Liskov's PBFT protocol achieved performance suitable for real-world replicated services~\cite{DBLP:conf/osdi/CastroL99}.
To be able to satisfy the high performance and security demands of global digital payments, such protocols often require further improvements and careful design~\cite{DBLP:journals/jsys/StathakopoulouDPV22,berger2023sok}.
Hereby, optimising desirable consensus properties, such as reaching low latency, i.\,e. \emph{finality} in a minimal number of communication rounds, may impose additional system assumptions~\cite{chou2025minimmitfastfinalityfaster} or present significant engineering challenges~\cite{xiang2025zaptos}.
Hence, any potential CBDC design must carefully consider the concrete choice of agreement mechanism and desired guarantees that it can offer. 

In Section~\ref{sssec:integrity} we discuss measures by which the integrity of global information is ensured and, in particular, distinguish between systems that guarantee \emph{total order} and \emph{no total order} of transactions.
While the former generally rely on consensus, the latter build upon a property that is observed in \cite{guerraoui2019consensus}, which shows that the \emph{consensus number}~\cite{DBLP:journals/toplas/Herlihy91} of an asset transfer object is 1. 
In essence, the approach leverages the fact that a global total order is not strictly necessary for concurrent, independent transactions in order to achieve desirable consistency guarantees for digital payment systems, i.\,e., prevent double-spending. 
Hence, more lightweight and round-efficient protocols than consensus, such as \emph{Byzantine Consistent Broadcast} can instead be used for asset-transfer, with several works~\cite{KKS22, SKK23, BZK24} building on this approach.
The drawback of this approach is that attempts to double spend can render the respective funds unusable. While this may serve as a deliberate disincentive, it can be difficult to explain to users in cases of erroneous double spending.

Conversely, work is being done to render systems with total order guarantees
more efficient, for example, Chop Chop~\cite{camaioni2024chop}, Sui~\cite{blackshear2024sui} and JUMBO~\cite{cheng2024jumbo}, which achieve 43 million and 65 million Tx/s, respectively, at peak performance.
These optimisations often involve multi-layered approaches, meaning a consensus-less fast path followed by a slower consensus path that establishes TO.

Regardless of the concrete choice of mechanism for ensuring consistency of the global state, considerations for the ability to apply future scalability measures should also be taken into account. 
One such approach is enabling a system to \emph{scale horizontally}, e.\,g. by distributing transaction processing and validation among shards. 

\section{Sharding and Concurrent Processing}
\label{xsec:sharding}
Horizontal scaling is another important approach to increasing system throughput.
In the context of digital payment systems, \emph{sharding} refers to partitioning the (global) state and transaction workload across multiple committees or processing units, such that disjoint subsets of transactions can be verified and executed independently, e.\,g.,~\cite{DM16,BDS20,TBA22,AMD23,LVF23}.
Hereby the concurrent nature of asset-transfer transactions can be leveraged~\cite{collins2020online,blackshear2024sui}.

The scalability benefits of sharding depend on the degree of \emph{transaction independence}, in the sense that transactions operate on disjoint portions of state and therefore commute, permitting them to be treated as independent concurrent objects~\cite{DBLP:journals/toplas/Herlihy91}.

From an architectural standpoint, negative information appears more readily shardable, e.\,g., by employing \emph{consistent hashing}~\cite{CGM21,TBA22}, however, designs such as Hamilton~\cite{LVF23} demonstrate that sharding can also be implemented on positive information.

Throughput can further benefit from \emph{concurrent execution} techniques, which exploit fine-grained independence between transactions.
Here, execution occurs optimistically, with conflicts detected and resolved during or after execution.
This approach has seen active development in high-performance distributed ledgers and smart contract platforms through techniques such as dependency-aware scheduling or speculative concurrency control~\cite{DBLP:conf/ppopp/GelashviliSXDLM23,blackshear2024sui,neiheiser2025anthemius}.
However, such methods still require the underlying workload to exhibit sufficient parallelism.

While sharding and optimistic execution both increase throughput by exploiting concurrency, they also introduce additional coordination challenges. 
In particular, ensuring \emph{cross-shard atomicity} requires protocols that guarantee consistent transaction outcomes across shards, and \emph{data availability} mechanisms are needed to ensure that each shard has access to the state required for validation. 
These requirements can impose additional communication overhead, meaning that the actual performance gains of horizontal scaling ultimately depend not only on workload independence, but also on the efficiency of the mechanisms that coordinate between shards~\cite{berger2023sok,DBLP:journals/access/YuWYNZL20,wang2019sok}.

These horizontal scaling approaches are generally intimately tied to the particular architecture and mechanisms for ensuring the integrity of global information, rendering direct comparisons difficult. 
A systematic framework for comparing these trade-offs in the context of CBDC architectures remains an important direction for future work.

\section{Realising Exact Amounts}
\label{xsec:exact_amounts}

The inability to pay exact amounts is solely a problem of designs building on Chaumian e-cash, i.\,e., Transfer Digital Coin patterns.  
Several solutions have been proposed to address this issue. 

\textbf{Cheques.} These allow arbitrary values to be spent by encoding the value in binary and spending only the parts of the cheque that represent a 1~\cite{CFN88}.  
First, though, they pose severe privacy risks, since for a refund of the remainder, the negative pattern of the spent coin is essentially revealed.  
Second, this approach may only work for a limited number of payments. For example, one cannot spend two uneven values with the same cheque.  

\textbf{Tree representation.} A coin is represented as a tree, where each node represents the sum of its child nodes. 
Spending a node means that none of its descendant or ancestor nodes can be spent~\cite{OO91, CFT98, CG10, CPS14}.  
Many values still require spending multiple nodes, and each node must be spent in a separate protocol.  

\textbf{Different denominations.} There are different signatures representing different denominations of value. 
This is the same concept that is used for cash. 
To be practical, this requires a system that implements a reasonable protocol to receive change or exchange coins quickly.  
Chaum et al.~\cite{CGM21} propose a change protocol, but it must be signed by a bank and involves an interactive protocol.  
Bauer et al.~\cite{BFQ21} argue that change in an e-cash system with offline functionality is more practical, but each coin still requires a separate payment protocol.  
Rial and Piotrowska~\cite{RP23} provide an in-depth discussion on the optimal set of denominations but do not offer an explicit change protocol.

\textbf{Coins of different values.} Coins of varying values can be withdrawn~\cite{ADE24}.  
This approach is still impractical because it has privacy risks.
If coins of arbitrary value can be withdrawn, the anonymity set for a given value may be very small, reducing sender anonymity.  
Additionally, the exact amount must be known in advance for the approach to be practical.

\addtolength{\textheight}{-5.1cm}
For designs based on Chaumian e-cash, many ideas have been proposed to pay exact amounts, but none are convincing in terms of usability so far.
As argued in~\cite{RP23}, when denominations are fixed, a user may have sufficient funds but still be unable to complete a transaction due to the lack of appropriate denominations.
This severely impacts usability.

In order for an e-cash system to be practical, two aspects are required:
\begin{enumerate}
    \item A compact method for spending multiple coins.
    \item A practical protocol for giving change (trivial in the case of offline payments).
\end{enumerate}

If the first aspect is highly efficient, the second can be neglected.  
In that case, only coins of the smallest denomination are used, and many of them are spent in a single payment.

\section{Offline Functionality}
\label{xsec:offline_discussion}

Offline functionality remains one of the biggest challenges in CBDC design, as it requires preventing or managing double-spending without constant online verification.
Designs offering offline functionality can be distinguished by their approach to either \emph{prevent} double-spending, \emph{detect-and-sanction} it retrospectively, or combine both strategies.

The \emph{detect-and-sanction} approach can be applied in early transferable e-cash systems, where a double-spender is identified once a coin is redeemed twice by different recipients with the bank.
The current design space suggests that the \emph{detect-and-sanction} approach is only possible when using negative global information. 
In this case, by cleverly choosing the way the nullifier is generated, a double-spender can be identified.

A limitation of the \emph{detect-and-sanction} approach is that the information size grows with each offline transfer~\cite{chaum92transferred}.
Two more recent designs move this growing amount of information to the users themselves, although this comes at the expense of privacy.
Fuchsbauer et al.~\cite{FPV09} achieve constant-size offline transfers, but at the cost of requiring every honest non-double-spender in the chain of offline transfers to prove their innocence in the event of a double-spend.
This, in turn, obliges all users to retain the necessary information to do so.
Similarly, PayOff~\cite{BZK24} offers offline functionality where users must iteratively request signatures for account states involved in offline payments when going back online.
Again, in the event of a double-spend, all honest users in the chain are required to prove their innocence, and for each offline payment, the user must store relevant information. 
Thus, total information still grows, but it is distributed among users rather than stored globally.
With today's hardware, growing transaction sizes are less of a problem than they were 30 years ago, although extreme cases in which users stay offline for extended periods might not be practical.

The \emph{detect-and-sanction} approach also raises the question of who bears responsibility in a double-spending incident: the recipient or the central bank.
Until the funds are recovered, one must cover the loss.
To preserve trust, it is suggested that the central bank be liable~\cite{BZK24}.
To reduce potential losses due to covering a double-spend, holding or spending limits~\cite{BZK24} are recommended.

Literature suggests that the \emph{prevent} double-spending approach requires trusted hardware (TH).
These designs have the drawback that, if the TH is compromised, the system may permit arbitrary double-spending in offline-payments.

That is why some designs combine the two approaches of handling double-spending.
These designs implement mechanisms that, in case the TH is compromised, can identify the double-spender in retrospect when settling online at a later point in time.

\begin{table}
    \caption{Design space of offline payment systems by double-spending handling method and limit implementation.}
    \centering
    \begin{tabular}{lccc}
        \toprule
        & \textbf{Detect-and-sanction} & \textbf{Prevent} & \textbf{Both} \\
        \midrule
        \textbf{Limits} &  & & \cite{BZK24} \\
        \textbf{No limits} & \makecell{\cite{OO91,CG08,FPV09}\\ \cite{BCF15,BFQ21}} & \cite{CGK20}$^{\mathrm{a}}$ & \cite{ADE24} \\
        \bottomrule
        \multicolumn{4}{p{\dimexpr0.6\linewidth-2\tabcolsep\relax}}{
            {\scriptsize $^{\mathrm{a}}$The paper by Christodorescu et al.~\cite{CGK20}, which adopts this approach, was not included in our comparison table due to the lack of a complete technical specification.}
        }
    \end{tabular}    
\label{tab:offline}
\end{table}
Table~\ref{tab:offline} categorises the analysed designs according to their approach to handling double-spending and whether they offer holding or spending limits.

In summary, while several strategies for offline functionality exist, each faces open challenges: complete reliance on trusted hardware, the deanonymisation of honest users in the case of double-spending, or growing transactions with each offline payment.
To date, no system providing offline functionality has been implemented.

\end{document}